\documentclass[preprintnumbers,amsmath,amssymb,12pt]{revtex4}

\newcommand{\ghza}{$\frac{1}{\sqrt{2}} \left[ \,\left| 000 \right\rangle_{123}
+ \left| 111 \right\rangle_{123}  \, \right]$}

\usepackage{graphicx}
\usepackage{dcolumn}
\usepackage{bm}
\begin{document}
\setlength{\topmargin}{0.2 in}
\baselineskip 24pt
\title{Multi-particle entanglement and generalized N-particle 
 teleportation using quantum statistical correlations}
\author{Atul Kumar} 
\author{Mangala Sunder Krishnan}%
\email{mangal@iitm.ac.in}
\affiliation{%
{Department of Chemistry, Indian Institute of
   Technology Madras, Chennai 600 036, India}}%
\date{\today}
\begin{abstract}
Construction of multi-particle entangled states and direct
teleportation of $N$-(spin 1/2) particles are important areas of quantum information
processing. A number of different schemes which have been presented
already, address the problem through controlled teleportation. In this article, a criterion based on standard
quantum statistical correlations employed in the many-body virial
expansion is used to determine maximum entanglement for a $N$-particle
state. These states
remain entangled through proper traces to states for a smaller number
of particles and can be generalized for arbitrary number of
particles. It is shown that they are quite useful in generalized,
$N$-particle, direct teleportation. The corresponding quantum gates
are also indicated for teleportation schemes from simple computational
basis states.

\end{abstract}
\maketitle
\begin{center}
\section*{I. INTRODUCTION}
\end{center}
Quantum teleportation deals with the transfer of information from
one remote location to another without physical transport of the information
or measurement on either side to confirm and verify the information
content [1]
. It rests on the quantum correlations for which there are
no classical analogues. Bennett and
coworkers proposed  a theoretical scheme in 1993 using the celebrated 
Bohm-Aharanov-Einstein-Podolsky-Rosen (EPR) pair and demonstrated the
transfer of two bits  of information 
without classical communication with a probability of 1/4 and with
classical communication (with speeds less than that of light in vacuum)
with probability 3/4 [2,3]. Transfer of the information content of a
spin 1/2 particle was thus shown in theory {\em with certainty} and with {\em no intermediate
observer monitoring/controlling} the process.
Many experiments [4-10] have been performed subsequently which provide
partial experimental support for this concept. A number of remarkable
theoretical concepts and schemes have also been invented for single
and multi-particle teleportation [11-18]. The original EPR pair which is
maximally entangled and played the role of the carrier of information
has been supplemented with less-than-maximally-entangled (with respect
to EPR pair) three and
four-particle correlated states [19-22]. \par
Some of the difficulties associated with multi-particle teleportation 
using correlated states are
\begin{enumerate}
\item  Maximum entanglement in the same sense as in the case of EPR pair
cannot be achieved under experimental conditions at present.
\item Three or four-particle
correlated states often require  additional measurement(s) (Charlie)
as an intermediate step as opposed to a pair of particles. 
\item The use of three-particle maximally entangled
Greenberger-Horne-Zeilinger (GHZ) states for the teleportation of a single
particle as quantum carrier and projection basis leads to failure of
the process even in theory since four out of eight GHZ states have
zero projections for the single-particle leading to null results. This
means that it will be impossible for the receiver to reconstruct the
unknown state sent by the sender [13].
\item Unitary transformations of the GHZ states which are robust with
respect to tracing of one of the particles and which obviate the need
for an intermediate observer have not been examined carefully. It must
be mentioned that Bell states are merely unitary transformations from
other two-particle states, which, however play a vital role in
information processing. They are different from an unentangled or
partially entangled state by a unitary transformation or two. Therefore it is necessary to identify proper,
genuinely entangled states for multiparticle teleportation which has not been
done in a systematic manner until now.
\item The extent of correlation between particles which is a direct measure
of entanglement is not defined uniformly for multiparticle systems, as
against a clear-cut definition for two particle systems [23-25]. \end{enumerate}
In this article we address all of the above problems and propose
schemes for both optics experiments and using quantum gates. This
article is organized as follows : \\
(a) In section II  different complete set(s) of orthonormal entangled
projection basis of three-particles is (are) proposed by using a unitary
transformation of GHZ basis. It is shown that in this new basis the observer Charlie is
not needed and direct teleportation results instead of controlled
process. Correlation coefficients are proposed as a measurement
criterion for entanglement of multiple paticles using standard
Ursell-Mayer type expansion based on the principles of many body
statistical mechanics. In addition to this, three-particle GHZ basis
has been used as {\em quantum carrier} as well as {\em projection basis} for single particle teleportation. \\
(b) In section III two different sets of genuinely entangled
four-particle states are proposed for the
teleportation of an arbitrary two-particle state. The
generalization of the same for $N$-particle system is also suggested in detail. \\
(c) In section IV teleportation using quantum gates and three and four
qubit computational bases is described with the appropriate quantum
circuit. This is followed by conclusion. 

\section*{II. THREE-PARTICLE ENTANGLED BASIS AND QUANTUM TELEPORTATION
  WITHOUT PARTICLE AVERAGING}
We propose a three-particle basis (123) in the following as
%
%
\begin{eqnarray}
\left| \chi \right\rangle_{123}^{(1),(2)}  = \frac{\left| \phi \rangle_{12}^+ \right. \otimes
\left| 0 \rangle_3 \right. \pm \left| \phi \rangle_{12}^- \right. \otimes
\left| 1 \rangle_3 \right.  }{\sqrt{2}} & , &
\left| \chi \right\rangle_{123}^{(3),(4)} = \frac{\left| \phi \rangle_{12}^+ \right. \otimes
\left| 1 \rangle_3 \right. \pm \left| \phi \rangle_{12}^- \right. \otimes
\left| 0 \rangle_3 \right. }{\sqrt{2}}\ , \nonumber \\
\left| \chi \right\rangle_{123}^{(5),(6)}  = \frac{\left| \psi \rangle_{12}^+ \right. \otimes
\left| 0 \rangle_3 \right. \pm \left| \psi\rangle_{12}^-  \right. \otimes
\left| 1 \rangle_3 \right.  }{\sqrt{2}} & {\rm and} &  \left| \chi \right\rangle_{123}^{(7),(8)} = \frac{\left| \psi \rangle_{12}^+\right. \otimes
\left| 1 \rangle_3 \right. \pm \left| \psi \rangle_{12}^- \right. \otimes
\left| 0 \rangle_3 \right. }{\sqrt{2}}
\end{eqnarray}
where
%
\begin{equation}
\left| \psi\right\rangle^{\pm}_{12}  = \frac{1}{\sqrt{2}} \left[ \,
\left| 01 \right\rangle_{12} \pm \left| 10 \right\rangle_{12} \,
\right]  ~~~{\rm and}~~~
\left| \phi\right\rangle^{\pm}_{12}  = \frac{1}{\sqrt{2}} \left[ \,
\left| 00 \right\rangle_{12} \pm \left| 11 \right\rangle_{12} \,
\right]\ .
 \end{equation}
These states are linear
combinations of three-particle GHZ states. The entanglement properties of these states are similar to the three-particle GHZ states if we
 consider the extent of correlation and to W states for robustness of entanglement with respect to  tracing of third particle [31,32]. The extent of correlation of entangled states  is measured with the help of the correlation
 coefficients defined using the well known statistical mechanical formula
 involving averages for many-body
 systems [26-30]. Correlation coefficients for two-particle
 and three-particle systems are defined as
\begin{eqnarray}
C^{ij}_{\alpha \beta } & = & \left\langle \sigma^i_{\alpha} \sigma^j_{\beta} \right\rangle
- \left\langle \sigma ^i_{\alpha} \right\rangle
 \left\langle \sigma ^j_{\beta} \right\rangle \  and \\
C^{ijk}_{\alpha \beta \gamma } & = & \left\langle \sigma^i_{\alpha} \sigma^j_{\beta} \sigma^k_{\gamma}
  \right\rangle
- \left\langle \sigma ^i_{\alpha} \right\rangle
 \left\langle \sigma ^j_{\beta} \sigma^k_{\gamma} \right\rangle
- \left\langle \sigma ^j_{\beta} \right\rangle
 \left\langle \sigma ^i_{\alpha} \sigma^k_{\gamma} \right\rangle
  - \left\langle \sigma ^k_{\gamma} \right\rangle
 \left\langle \sigma ^i_{\alpha} \sigma^j_{\beta} \right\rangle
+ 2\left\langle \sigma ^i_{\alpha} \right\rangle
\left\langle \sigma ^j_{\beta} \right\rangle
\left\langle \sigma ^k_{\gamma} \right\rangle . \nonumber \\ & & \end{eqnarray}
%
%
where $\sigma$'s are the Pauli
spin matrices for the indicated particles and
\{ $\alpha$, $\beta$, $\gamma$ = ${x , y , z}$\}. Table I  lists the
non-zero correlation coefficients for the four Bell states of a pair of
particles.
The averages are calculated and compared, for the states \{$\left| \chi
\right\rangle^{(1)} _{123}  - \left| \chi \right\rangle^{(8)}_{123}$\}
and the eight GHZ states in Table II .
The extent of correlation between three particles is the same in both the
sets as can be seen from Table II; however, the advantage associated
with the entangled basis proposed here over GHZ states is that the states in Eq. 1
are robust with respect to the  tracing of 3rd particle, i.e., on tracing the remaining 
pair is still entangled [31-33]. It is worth
mentioning here that a linear combination of three-particle GHZ states
such as
\begin{equation} 
\left| \chi\right\rangle_{123} =
\frac{1}{2}\left[\,\left|000\right\rangle_{123} + \left|
    110\right\rangle_{123} + \left| 001\right\rangle_{123} + \left|
    111\right\rangle_{123} \right]
\end{equation} 
 possesses no genuine three-particle quantum
correlation. When calculated, all the correlation coefficients
associated with the above state shows zero value, because the state can be expressed as the direct product state
$
\left| \chi \right\rangle _{123}=\frac{1}{\sqrt{2}}
[\left| 00 \right\rangle _{12} + \left| 11 \right\rangle _{12}]
\otimes\frac{1}{\sqrt{2}}[\left| 0 \right\rangle _{3} + \left| 1 \right\rangle _{3}].
$

If we use GHZ basis as a quantum channel to
teleport unknown information encoded in a single particle then the basis
set given by Eq. 1  can be used as projection basis and it obviates the
earlier difficulties of missing elements of basis set (when GHZ basis
functions are used as a projection basis). \par 

In this scheme, the single-particle information is with Alice, given by
$ \left| \phi \right\rangle_1 = a \left| 0 \right\rangle_1  + b \left|
  1 \right\rangle_1 $
where $ \left\langle \phi _1 | \phi _1 \right\rangle = 1$ and
$|a|^2 +|b|^2 = 1$.
The GHZ state (234), $ \left| \psi  \right\rangle^{GHZ} _{234} =
\frac{1}{\sqrt{2}} \left[ \,\left| 000 \right\rangle _{234} +
 \left| 111 \right\rangle _{234} \right] $, shared by Alice (23)
and Bob (4) along with particle (1), gives rise to product representation
for the four particle state as
%
%
\begin{equation}
\left| \psi \right\rangle _{1234} = \left| \phi \right\rangle _1 \otimes
\left| \psi \right\rangle ^{GHZ}_{234} = \frac{a \left[ \, \left| 0000
\right\rangle _{1234}  + \left| 0111 \right\rangle _{1234} \, \right] }{\sqrt{2}} +
\frac{b \left[ \, \left| 1000 \right\rangle _{1234} + \left| 1111
\right\rangle _{1234}  \, \right] }{\sqrt{2}}.
\end{equation}
%
%
Alice's measurement process using her basis states 
\{$\left| \chi
\right\rangle^{(1)} _{123}  - \left| \chi \right\rangle^{(8)}_{123}$\} 
is
based on the following decomposition of her state,
%
%
\begin{eqnarray}
\left| \psi \right\rangle _{1234} & = &
\frac{1}{2\sqrt{2}} \left\{  \left| \chi  \right\rangle^{(1)} _{123} \left[
a\left| 0 \right\rangle _4 - b \left| 1 \right\rangle _4 \right] +
\left| \chi  \right\rangle^{(2)} _{123} \left[
a\left| 0 \right\rangle _4 + b \left| 1 \right\rangle _4 \right] +
\left| \chi \right\rangle ^{(3)}_{123} \left[
a\left| 0 \right\rangle _4 + b \left| 1 \right\rangle _4 \right]
\right. \nonumber \\ & &
+\left| \chi  \right\rangle^{(4)} _{123} \left[
-a\left| 0 \right\rangle _4 + b \left| 1 \right\rangle _4 \right] +
\left| \chi  \right\rangle^{(5)} _{123} \left[
a\left| 1 \right\rangle _4 + b \left| 0 \right\rangle _4 \right] +
\left| \chi  \right\rangle^{(6)} _{123} \left[
-a\left| 1 \right\rangle _4 + b \left| 0 \right\rangle _4 \right] \nonumber \\ & &
+\left. \left| \chi  \right\rangle^{(7)} _{123} \left[
a\left| 1 \right\rangle _4 - b \left| 0 \right\rangle _4 \right] +
\left| \chi  \right\rangle^{(8)}_{123} \left[
a\left| 1 \right\rangle _4 + b \left| 0 \right\rangle _4 \right] \right\}.
\end{eqnarray}
%
%
\newcommand{\chb}{$\left| \chi_{123}^{(5)} \right\rangle $}
Four equally probable results for Bob's
particle (4) are possible and also there is no need for another observer to
assist Alice in transferring the information to Bob. One of the four
probable outcomes is a direct teleportation as indicated by $I^{4}$
in Table III. The other three outcomes require one unitary transformation as given in
Table III.  It is important to mention here that our protocol is an
example of {\em  direct teleportation as against controlled teleportation} where
Alice needs Charlie to assist her in sending the unknown information
to Bob with Charlie carrying one of the entangled particles out of
three. In the present case Alice has two particles on her side so that she
can do a direct three-particle measurement. Our protocol also
overcomes earlier difficulties such as getting null results in four out of
eight projections performed by Alice so that the receiver would never be
able to recover the unknown state in four such cases.\par

Gorbachev and Trubilko [13] have shown that, by using a basis which consists
of the direct product of a single-particle state with a two-particle
entangled state, the teleportation of an arbitrary EPR pair can be realized. In
their scheme, Alice interacts her unknown EPR state with the GHZ state and
projects her three particles on the three-particle basis states (123),
given by ${(\left|\pi\right\rangle^{\pm}_{1}\
\otimes\left|\psi\right\rangle^{\pm}_{23} , \left|\pi\right\rangle^{\pm}_{1}\
\otimes\left|\phi\right\rangle^{\pm}_{23})}$  where $\left| \psi\right\rangle^{\pm}_{23}$ and $\left| \phi\right\rangle^{\pm}_{23}$
are Bell States for the pair (23) ( Eq. 1) and  $\left| \pi  \right\rangle^{\pm}_{1} =
\frac{1}{\sqrt{2}} \left[ \,\left| 0 \right\rangle _{1} {\pm}
 \left| 1 \right\rangle _{1} \right]$ . This
leads to Bob's EPR pair being in a teleported state and the original
arbitrary EPR state can be found by doing a simple unitary transformation for
all the outcomes.
We have demonstrated above that the projection basis in Eq. 1 can be
used for satisfactory teleportation of a single particle. To
contrast with the result of Gorbachev and Trubilko we give
below the scheme to teleport an arbitrary EPR pair through a different set of
three-particle entangled projection basis given by
%
\begin{eqnarray}
\left| \varphi \right\rangle ^{(1),(2)} _{123} =
\frac{ \left| 0 \right\rangle _1 \otimes \left| \phi \right\rangle ^+_{23}
\pm  \left| 1 \right\rangle _1 \otimes \left| \phi \right\rangle
^-_{23}}{\sqrt{2}}  & , &
\left| \varphi \right\rangle ^{(3),(4)} _{123} =
\frac{ \left| 1 \right\rangle _1 \otimes \left| \phi \right\rangle ^+_{23}
\pm  \left| 0 \right\rangle _1 \otimes \left| \phi \right\rangle
^-_{23}}{\sqrt{2}}\ ,
\nonumber \\
\left| \varphi \right\rangle ^{(5),(6)} _{123} =
\frac{ \left| 0 \right\rangle _1 \otimes \left| \psi \right\rangle ^+_{23}
\pm  \left| 1 \right\rangle _1 \otimes \left| \psi \right\rangle
^-_{23}}{\sqrt{2}} & {\rm and} &
\left| \varphi \right\rangle ^{(7),(8)} _{123} =
\frac{ \left| 1 \right\rangle _1 \otimes \left| \psi \right\rangle ^+_{23}
\pm  \left| 0 \right\rangle _1 \otimes \left| \psi \right\rangle ^-_{23}}{\sqrt{2}}.
\end{eqnarray}
%
The sets given by Eq. 1 and Eq. 8 differ in the ordering of the
particles. In angular momentum algebraic parlance they refer to
different coupling schemes and are related to each other through 
a unitary tranformation containing a 6j coefficient. Alice can use any one of
the Bell pairs to transfer
information to Bob for e.g. $ \left| \psi \right\rangle _{12} =  a \left| 01 \right\rangle_{12}  + b \left|
10 \right\rangle_{12}$. The GHZ
state $ \left| \psi  \right\rangle^{GHZ} _{345} =
\frac{1}{\sqrt{2}} \left[ \,\left| 000 \right\rangle _{345} +
\left| 111 \right\rangle _{345} \right]$ is composed of Alice's particle 3 and Bob's particles
4 and 5. Thus the unknown two-particle state and the GHZ state give
rise to a five-particle state as
%
\begin{equation}
\left| \psi \right\rangle _{12345}= \left| \psi \right\rangle _{12} \otimes
\left| \psi \right\rangle ^{GHZ}_{345} =
[ a \left| 01 \right\rangle _{12} + b \left| 10 \right\rangle _{12}]
\otimes \left[\frac{\left| 000 \right\rangle _{345} + \left| 111 \right\rangle _{345}}{\sqrt{2}} \right].
\end{equation}
%
Alice's measurements are projections on the three-particle (123)
states given by Eq. 8, namely,
%
\begin{eqnarray}
\lefteqn{\left|\psi\right\rangle_{12345} =} & & \nonumber \\ & & 
\frac{1}{2\sqrt{2}} \left\{ \left| \varphi \right\rangle ^{(1)}_{123} \left[
a \left| 11 \right\rangle _{45} + b \left| 00 \right\rangle _{45} \right]
+ \left| \varphi \right\rangle ^{(2)}_{123} \left[
a \left| 11 \right\rangle _{45} - b \left| 00 \right\rangle _{45} \right]
+ \left| \varphi \right\rangle ^{(3)}_{123} \left[
-a \left| 11 \right\rangle _{45} + b \left| 00 \right\rangle _{45} \right]
\right.  \nonumber \\ & &
+ \left| \varphi \right\rangle ^{(4)}_{123} \left[
a \left| 11 \right\rangle _{45} + b \left| 00 \right\rangle _{45} \right]
+ \left| \varphi \right\rangle ^{(5)}_{123} \left[
a \left| 00 \right\rangle _{45} + b \left| 11 \right\rangle _{45} \right]
+ \left| \varphi \right\rangle ^{(6)}_{123} \left[
a \left| 00 \right\rangle _{45} - b \left| 11 \right\rangle _{45} \right]
\nonumber \\ & &
\left. + \left| \varphi \right\rangle ^{(7)}_{123} \left[
-a \left| 00 \right\rangle _{45} + b \left| 11 \right\rangle _{45} \right]
+ \left| \varphi \right\rangle ^{(8)}_{123} \left[
a \left| 00 \right\rangle _{45} + b \left| 11 \right\rangle _{45} \right]
\right\}.
\end{eqnarray}
%
The two-particle (45) state is in one of the four distinct outcomes which can be
transformed back to the original arbitrary EPR pair easily through a single
qubit unitary transformation by Bob. The required transformations are listed in
Table IV. \par
The basis sets given by Eq. 8 and Eq. 1  work successfully as projection
bases for the teleportation of a single particle and an arbitrary EPR pair
respectively as well, by doing two different two-qubit transformations on Alice's side.
For example, if we consider teleportation of a particle through
GHZ state using projection basis given by Eq. 8, the direct product state
of four particles is
%
\begin{equation}
\left| \psi  \right\rangle _{1234} = \frac{a}{\sqrt{2}} \left[ \,
  \left| 0000 \right\rangle _{1234} + \left| 0111 \right\rangle _{1234} \right] +
\frac{b}{\sqrt{2}} \left[ \, \left| 1000 \right\rangle _{1234} + \left| 1111 \right\rangle _{1234} \right].
\end{equation}
%
A direct three-particle measurement using basis set of Eq. 8 will not
be able to achieve the desired task, thus Alice does two different
unitary transformations on her particles (23) and (12) respectively given by
unitary matrices $U^{(1)}_{23}$ and $U^{(2)}_{12}$ , where
%
\begin{equation} U_{23}^{(1)} = \left( \begin{array}{cccc}
1 & 0 & 0 & 0 \\ 0 & 1 & 0 & 0 \\ 0 & 0 & 0 & 1 \\ 0 & 0 & 1 & 0 \end{array}
\right) _{23} ~~~~ {\rm and} ~~~~
U_{12}^{(2)} = \left( \begin{array}{cccc}
1 & 0 & 0 & 0 \\ 0 & 1 & 0 & 0 \\ 0 & 0 & 0 & 1 \\ 0 & 0 & 1 & 0 \end{array}
\right) _{12}.  \end{equation}
%
The two transformations on $\left|\psi\right\rangle_{1234}$ are represented as
\begin{equation}
\left| \psi \right\rangle ^{(1)}_{1234} = I_1 \otimes U_{23}^{(1)} \otimes I_4
\left| \psi \right\rangle _{1234} =
\frac{a}{\sqrt{2}} \left[ \, \left| 0000 \right\rangle _{1234} +
\left| 0101 \right\rangle _{1234} \right] +
\frac{b}{\sqrt{2}} \left[ \, \left| 1000 \right\rangle _{1234} +
\left| 1101 \right\rangle _{1234} \right] \end{equation} 
and
%
\begin{equation}
\left| \psi \right\rangle ^{(2)}_{1234} = U_{12}^{(2)} \otimes I_3 \otimes I_4
\left| \psi \right\rangle ^{(1)} _{1234} =
\frac{a}{\sqrt{2}} \left[ \, \left| 0000 \right\rangle _{1234} +
\left| 0101 \right\rangle _{123} \right] +
\frac{b}{\sqrt{2}} \left[ \, \left| 1100 \right\rangle _{1234} +
\left| 1001 \right\rangle _{1234} \right] \end{equation}
%
where $I_{i}$ represents an identity matrix for $i^{th}$ particle.
$\left|\psi\right\rangle^{(2)}_{1234}$ is re-expressed using the basis set of Eq. 8 for Alice to
project her particles onto any of these eight states, as
%
\begin{eqnarray}
\left| \psi \right\rangle ^{(2)}_{1234} & = &
\frac{1}{2\sqrt{2}} \left\{ \left| \varphi \right\rangle ^{(1)}_{123} \left[
a \left| 0 \right\rangle _4 + b \left| 1 \right\rangle _4 \right]
+ \left| \varphi \right\rangle ^{(2)}_{123} \left[
a \left| 0 \right\rangle _4 - b \left| 1 \right\rangle _4 \right]
+ \left| \varphi \right\rangle ^{(3)}_{123} \left[
a \left| 0 \right\rangle _4 + b \left| 1 \right\rangle _4 \right]
\right.  \nonumber \\ & &
+ \left| \varphi \right\rangle ^{(4)}_{123} \left[
-a \left| 0 \right\rangle _4 + b \left| 1 \right\rangle _4 \right]
+ \left| \varphi \right\rangle ^{(5)}_{123} \left[
a \left| 1 \right\rangle _4 - b \left| 0 \right\rangle _4 \right]
+ \left| \varphi \right\rangle ^{(6)}_{123} \left[
a \left| 1 \right\rangle _4 + b \left| 0 \right\rangle _4 \right]
\nonumber \\ & &
\left. + \left| \varphi \right\rangle ^{(7)}_{123} \left[
-a \left| 1 \right\rangle _4 + b \left| 0 \right\rangle _4 \right]
+ \left| \varphi \right\rangle ^{(8)}_{123} \left[
a \left| 1 \right\rangle _4 + b \left| 0 \right\rangle _4 \right]
\right\}.
\end{eqnarray}
%
Depending on Alice's measurement Bob may have to apply appropriate
unitary transformations to recover Alice's information. \par
It is worth mentioning here that any of the projection basis given by
Eq. 8 can be used as a quantum {\em carrier} to teleport a single particle
using GHZ basis as the {\em projection basis} for Alice's qubits. The basis
set in Eq. 8 possesses same correlation coefficients as that of GHZ
states with respect to particle tracing, and is more robust as compared to GHZ states, which makes it a
suitable set for being a quantum carrier.
The four-particle direct product state composed of Alice's information
and the quantum carrier $\left| \varphi \right\rangle ^{(1)} _{234}$ is
%
\begin{eqnarray}
\left| \psi \right\rangle _{1234} & = &  \left[ \,a  \left| 0 \right\rangle _1
+ b \left| 1 \right\rangle _1 \right] \otimes \left| \varphi \right\rangle ^{(1)} _{234}  \nonumber \\ & = &
\frac{a}{2} \left[ \, \left| 0000 \right\rangle _{1234} +
\left| 0011 \right\rangle _{1234} +
\left| 0100 \right\rangle _{1234} -
\left| 0111 \right\rangle _{1234} \right] \nonumber \\ & = &
+\frac{b}{2} \left[ \, \left| 1000 \right\rangle _{1234} +
\left| 1011 \right\rangle _{1234} +
\left| 1100 \right\rangle _{1234} -
\left| 1111 \right\rangle _{1234} \right]. \end{eqnarray}
%
A simple decomposition based on Alice's measurement in the GHZ basis (123) leads to
%
\begin{eqnarray}
\lefteqn{\left| \psi \right\rangle _{1234}  = } & & \nonumber \\ & & 
\frac{1}{2\sqrt{2}}\left\{
\left[ \, \left| 000 \right\rangle _{123} + \left| 111 \right\rangle _{123}  \right] \left[ a \left| 0 \right\rangle _4 -b \left| 1 \right\rangle _4 \right] +
\left[ \, \left| 000 \right\rangle _{123} - \left| 111 \right\rangle
_{123}  \right] \left[ a \left| 0 \right\rangle _4 +b \left| 1
\right\rangle _4 \right] \right. \nonumber \\  &  & +
\left[ \, \left| 001 \right\rangle _{123} + \left| 110 \right\rangle _{123}  \right] \left[ a \left| 1 \right\rangle _4 +b \left| 0 \right\rangle _4 \right] +
\left[ \, \left| 001 \right\rangle _{123} - \left| 110 \right\rangle
_{123}  \right] \left[ a \left| 1 \right\rangle _4 -b \left| 0
\right\rangle _4 \right] \nonumber \\ & & +
\left[ \, \left| 010 \right\rangle _{123} + \left| 101 \right\rangle _{123}  \right] \left[ a \left| 0 \right\rangle _4 +b \left| 1 \right\rangle _4 \right] +
\left[ \, \left| 010 \right\rangle _{123} - \left| 101 \right\rangle
_{123}  \right] \left[ a \left| 0 \right\rangle _4 -b \left| 1
\right\rangle _4 \right] \nonumber \\ &  & +
\left.
\left[ \, \left| 011 \right\rangle _{123} + \left| 100 \right\rangle _{123}  \right] \left[ -a \left| 1 \right\rangle _4 +b \left| 0 \right\rangle _4 \right] +
\left[ \, \left| 011 \right\rangle _{123} - \left| 100 \right\rangle
_{123}  \right] \left[ -a \left| 1 \right\rangle _4 -b \left| 0
\right\rangle _4 \right] \right\}. \nonumber \\ & & \end{eqnarray}
The above expression shows direct teleportation of Alice's information in
two of the cases with equal probabilities. In all other six cases appropriate unitary transformations
are needed subject to Alice's measurement results.
In addition to this we would like to mention another but similar set
of three-particle states having the same degree of correlation and
robustness as states given by Eq. 1 and Eq. 8. These states are  used
as {\em projection basis} for the
teleportation of a particle and an arbitrary EPR pair through
three-particle GHZ state and can be used as a {\em quantum carrier} for
single-particle teleportation using three-particle GHZ basis as
projections. They are given as 
\begin{eqnarray}
\left| \chi \right\rangle_{123}^{(1)',(2)'}  = \frac{\left| \phi \rangle_{13}^+ \right. \otimes
\left| 0 \rangle_2 \right. \pm \left| \phi \rangle_{13}^- \right. \otimes
\left| 1 \rangle_2 \right.  }{\sqrt{2}} & , &
\left| \chi \right\rangle_{123}^{(3)',(4)'} = \frac{\left| \phi \rangle_{13}^+ \right. \otimes
\left| 1 \rangle_2 \right. \pm \left| \phi \rangle_{13}^- \right. \otimes
\left| 0 \rangle_2 \right. }{\sqrt{2}}\ , \nonumber \\
\left| \chi \right\rangle_{123}^{(5)',(6)'}  = \frac{\left| \psi \rangle_{13}^+ \right. \otimes
\left| 0 \rangle_2 \right. \pm \left| \psi\rangle_{13}^-  \right. \otimes
\left| 1 \rangle_2 \right.  }{\sqrt{2}} & {\rm and} &  \left| \chi \right\rangle_{123}^{(7)',(8)'} = \frac{\left| \psi \rangle_{13}^+\right. \otimes
\left| 1 \rangle_2 \right. \pm \left| \psi \rangle_{13}^- \right. \otimes
\left| 0 \rangle_2 \right. }{\sqrt{2}} . \nonumber \\ & & 
\end{eqnarray}
Thus we demonstrate that for the teleportation of a particle, GHZ basis functions
can be used as {\em projections} as
well as {\em quantum carriers}. 

\section*{ III. MULTIPARTITE ENTANGLEMENT AND \\ TELEPORTATION}
In this section the entanglement properties of four-
particle entangled states and generalization of the same to
multi-particle systems are given. 
In addition, we discuss the teleportation of arbitrary
two-particle  state with generalization of the protocol
to a  $N$-particle system
using genuine multipartite states as quantum channels.
The extent of entanglement is assessed by a well established statistical mechanical formula for correlation
coefficients [28-30]. Correlation measures for more than three particles can be
defined using Ursell-Mayer type cluster coefficients. The use of
quantum virial coefficients [29-30] as a criterion for determining correlations
between spins (particles) resolves ambiguities in defining
the degree of entanglement  between multiple particles as against a clear existing 
definition for a pair of particles.
We therefore, use the following expression for the four particle correlation coefficient,
\begin{eqnarray} 
C^{1234}_{\alpha \beta \gamma \delta} & = & 
\left\langle \sigma^1_{\alpha} \sigma^2_{\beta} \sigma^3_{\gamma}\sigma^4_{\delta}
  \right\rangle
- \left\langle \sigma ^1_{\alpha} \right\rangle
 \left\langle \sigma ^2_{\beta} \sigma^3_{\gamma}
   \sigma^4_{\delta}\right\rangle
- \left\langle \sigma ^2_{\beta} \right\rangle
 \left\langle \sigma ^1_{\alpha} \sigma^3_{\gamma}
   \sigma^4_{\delta}\right\rangle 
- \left\langle \sigma ^3_{\gamma} \right\rangle
 \left\langle \sigma ^1_{\alpha} \sigma^2_{\beta}
   \sigma^4_{\delta}\right\rangle \nonumber \\ & - &
 \left\langle \sigma ^4_{\delta} \right\rangle
 \left\langle \sigma ^1_{\alpha} \sigma^2_{\beta}
   \sigma^3_{\gamma}\right\rangle 
+2 \left\langle \sigma ^1_{\alpha} \right\rangle \left\langle \sigma ^2_{\beta} \right\rangle
 \left\langle \sigma ^3_{\gamma} \sigma^4_{\delta} \right\rangle 
+2 \left\langle \sigma ^1_{\alpha} \right\rangle \left\langle \sigma ^3_{\gamma} \right\rangle
 \left\langle \sigma ^2_{\beta} \sigma^4_{\delta} \right\rangle 
+2 \left\langle \sigma ^1_{\alpha} \right\rangle \left\langle \sigma ^4_{\delta} \right\rangle
 \left\langle \sigma ^2_{\beta} \sigma^3_{\gamma} \right\rangle
 \nonumber \\ & +2 &
\left\langle \sigma ^2_{\beta} \right\rangle \left\langle \sigma ^3_{\gamma} \right\rangle
 \left\langle \sigma ^1_{\alpha} \sigma^4_{\delta} \right\rangle 
+ 2 \left\langle \sigma ^2_{\beta} \right\rangle \left\langle \sigma ^4_{\delta} \right\rangle
 \left\langle \sigma ^1_{\alpha} \sigma^3_{\gamma} \right\rangle
+2 \left\langle \sigma ^3_{\gamma} \right\rangle \left\langle \sigma ^4_{\delta} \right\rangle
 \left\langle \sigma ^1_{\alpha} \sigma^2_{\beta} \right\rangle  
- \left \langle \sigma^1_{\alpha} \sigma^2_{\beta} \right \rangle 
\left \langle \sigma^3_{\gamma} \sigma^4_{\delta} \right \rangle \nonumber \\ & -& 
 \left \langle \sigma^1_{\alpha} \sigma^3_{\gamma} \right \rangle 
\left \langle \sigma^2_{\beta} \sigma^4_{\delta} \right \rangle 
- \left \langle \sigma^1_{\alpha} \sigma^4_{\delta} \right \rangle 
\left \langle \sigma^2_{\beta} \sigma^3_{\gamma} \right \rangle 
- 6 \left \langle \sigma^1_{\alpha} \right \rangle \left \langle \sigma^2_{\beta} \right \rangle 
\left \langle \sigma^3_{\gamma} \right \rangle \left \langle \sigma^4_{\delta} \right \rangle .
\end{eqnarray}
The general expression for the $N$-particle correlation coefficient can be obtained 
by solving the equations for cluster functions derived formally from
 the $N$-th quantum virial coefficient (quantum trace) [28-29]. \par
Correlation coefficients calculated by using the above expression have been
used as entanglement criterion for four-particle states in this paper. Rigolin proposed a set of
generalized Bell basis set for the teleportation of two-particle state
which is a direct product of two, two-particle
Bell states. The four-particle correlation coefficients for all these 16 orthonormal
generalized Bell basis are zero.
One can form another orthonormal basis set of four-particle states similar to Rigolin's
generalized Bell basis, one member of which can be given as
\begin{equation}
\left| \psi\right\rangle_{1234}^{(1)} =
\frac{1}{2}\left[\,\left|0000\right\rangle_{1234}+ \left|
    1001\right\rangle_{1234}+\left| 0110\right\rangle_{1234}+\left|
    1111\right\rangle_{1234} \right]
\end{equation}
This set also works properly for the teleportation of arbitrary
two-particle states with only single qubit unitary transformations on
Bob's side. However, it also does not
have genuine four particle entanglement (all the four-particle correlation
coeficients are zero). Rigolin's state(s) and the basis shown above
(Eq. 20) are linear combinations of
GHZ states and possess no genuine multi-particle correlation which is
quite similar to what we have discussed in the previous section with
three-particle states (Eq. 5). An example of a set of sixteen
four-particle states which possess genuine four-particle correlation
(Eq. 19) is given by Yeo and Chua [18]. The non-zero correlation
coefficients associated with their set of states is indicated in
Table V. \par 
Here we propose two different sets of states which also possess genuine
four-particle entanglement which can be used successfully for the
teleportation of arbitrary two-particle states.

\subsection{\label{sec:level2}First set}

In the first scheme, we propose a set of orthonormal basis states which are linear
combinations of GHZ states, namely
\begin{eqnarray}
\left| \phi \right\rangle_{1234}^{(1),(2)} & =  & \frac{\left| 0 \rangle_1
  \right. \otimes \left| \phi \rangle_{23}^+ \right. \otimes
\left| 0 \rangle_4 \right. \pm \left| 1 \rangle_1
  \right. \otimes \left| \phi \rangle_{23}^- \right. \otimes
\left| 1 \rangle_4 \right.  }{\sqrt{2}} ,  \nonumber \\ 
\left| \phi \right\rangle_{1234}^{(3),(4)} & = & \frac{\left| 0 \rangle_1
  \right. \otimes \left| \phi \rangle_{23}^+ \right. \otimes
\left| 1 \rangle_4 \right. \pm \left| 1 \rangle_1
  \right. \otimes \left| \phi \rangle_{23}^- \right. \otimes
\left| 0 \rangle_4 \right.  }{\sqrt{2}} , \nonumber \\
\left| \phi \right\rangle_{1234}^{(5),(6)} & =  & \frac{\left| 0 \rangle_1
  \right. \otimes \left| \psi \rangle_{23}^+ \right. \otimes
\left| 0 \rangle_4 \right. \pm \left| 1 \rangle_1
  \right. \otimes \left| \psi \rangle_{23}^- \right. \otimes
\left| 1 \rangle_4 \right.  }{\sqrt{2}}, \nonumber \\ 
\left| \phi \right\rangle_{1234}^{(7),(8)} & = & \frac{\left| 0 \rangle_1
  \right. \otimes \left| \psi \rangle_{23}^+ \right. \otimes
\left| 1 \rangle_4 \right. \pm \left| 1 \rangle_1
  \right. \otimes \left| \psi \rangle_{23}^- \right. \otimes
\left| 0 \rangle_4 \right.  }{\sqrt{2}},  \nonumber \\
\left| \phi \right\rangle_{1234}^{(9),(10)}& =  & \frac{\left| 1 \rangle_1
  \right. \otimes \left| \phi \rangle_{23}^+ \right. \otimes
\left| 0 \rangle_4 \right. \pm \left|0 \rangle_1
  \right. \otimes \left| \phi \rangle_{23}^- \right. \otimes
\left| 1 \rangle_4 \right.  }{\sqrt{2}}, \nonumber \\ 
\left| \phi \right\rangle_{1234}^{(11),(12)} & =  & \frac{\left| 1 \rangle_1
  \right. \otimes \left| \phi \rangle_{23}^+ \right. \otimes
\left| 1 \rangle_4 \right. \pm \left| 0 \rangle_1
  \right. \otimes \left| \phi \rangle_{23}^- \right. \otimes
\left| 0 \rangle_4 \right.  }{\sqrt{2}} , \nonumber \\
\left| \phi \right\rangle_{1234}^{(13),14)} & =  & \frac{\left| 1 \rangle_1
  \right. \otimes \left| \psi \rangle_{23}^+ \right. \otimes
\left| 0 \rangle_4 \right. \pm \left| 0 \rangle_1
  \right. \otimes \left| \psi \rangle_{23}^- \right. \otimes
\left| 1 \rangle_4 \right.  }{\sqrt{2}} {~~~\rm and} \nonumber \\
\left| \phi \right\rangle_{1234}^{(15),(16)} & =  & \frac{\left| 1 \rangle_1
  \right. \otimes \left| \psi \rangle_{23}^+ \right. \otimes
\left| 1 \rangle_4 \right. \pm \left| 0 \rangle_1
  \right. \otimes \left| \psi \rangle_{23}^- \right. \otimes
\left| 0 \rangle_4 \right.  }{\sqrt{2}}.
\end{eqnarray}

The above basis has the advantage that it has  genuine four-particle entanglement as these states cannot be written as the direct
product of two-particle states. We have calculated the non-zero correlation
coefficients for this set and listed them in Table VI. The maximum value $({\pm}1)$ of
correlation coefficients suggests that the extent of entanglement is maximum
between four-particles which supports its use as a quantum carrier as
well as a projection basis. These states are robust with respect to two
particle tracing (14, 23) such that when traced over (14) or (23), the other two
particles will be in a correlated state. \par
Teleportation protocol for the two
particle state $\left|\phi\right\rangle_{12} =
a\left|00\right\rangle_{12}+b\left|01\right\rangle_{12}
+c\left|10\right\rangle_{12}+d\left|11\right\rangle_{12}$
is given below. Alice and Bob can use any one of the entangled state
as a quantum carrier described in the given set, e.g.,
\begin{equation}
\left| \phi\right\rangle_{3456}^{(1)} =
\frac{1}{2}\left[\,\left|0000\right\rangle_{3456}+ \left|
    1001\right\rangle_{3456}+\left| 0110\right\rangle_{3456}-\left|
    1111\right\rangle_{3456} \right]
\end{equation}
where particles 3 and 4 are with Alice and 5 and 6 are with Bob.
 The initial direct product of the six-particle state is given by
\begin{eqnarray}
\left| \psi \right\rangle_{123456} & = & \frac{a}{2}
\left[\left|
    000000 \right\rangle _{123456} +  \left|
    001001 \right\rangle _{123456} + \left|
    000110 \right\rangle _{123456} -  \left|
    001111 \right\rangle _{123456}  \right] \nonumber \\ & + & \frac{b}{2}\left[ \left|
    010000 \right\rangle _{123456} +  \left|
    011001 \right\rangle _{123456} + \left|
    010110 \right\rangle _{123456} -  \left|
    011111 \right\rangle _{123456}  \right] \nonumber \\ & + &
\frac{c}{2}
\left[\left|
    100000 \right\rangle _{123456} +  \left|
    101001 \right\rangle _{123456} + \left|
    100110 \right\rangle _{123456} -  \left|
    101111 \right\rangle _{123456}  \right] \nonumber \\ & +  & \frac{d}{2}\left[ \left|
    110000 \right\rangle _{123456} +  \left|
    111001 \right\rangle _{123456} + \left|
    110110 \right\rangle _{123456} -  \left|
    111111 \right\rangle _{123456}  \right].
\end{eqnarray}
Direct teleportation is the result, as is seen when Eq. 23 is re-expressed in the
basis set of Eq. 21, i.e.
\begin{eqnarray}
\left|\psi\right\rangle_{123456} & =  &
\frac{\left|\phi\right\rangle^{(1)}_{1234}}{4}\left[a\left|00\right\rangle_{56}+b\left|01\right\rangle_{56}+c\left|10\right\rangle_{56}+d\left|11\right\rangle_{56}\right]
\nonumber \\
 &  +  & \frac{\left|\phi\right\rangle^{(2)}_{1234}}{4}\left[a\left|00\right\rangle_{56}+b\left|01\right\rangle_{56}-c\left|10\right\rangle_{56}-d\left|11\right\rangle_{56}\right]
\nonumber \\
& + & \frac{\left|\phi\right\rangle^{(3)}_{1234}}{4}\left[a\left|10\right\rangle_{56}-b\left|11\right\rangle_{56}+c\left|00\right\rangle_{56}-d\left|01\right\rangle_{56}\right]
\nonumber \\
& + &\frac{\left|\phi\right\rangle^{(4)}_{1234}}{4}\left[a\left|10\right\rangle_{56}-b\left|11\right\rangle_{56}-c\left|00\right\rangle_{56}+d\left|01\right\rangle_{56}\right]
\nonumber \\
& + & \frac{\left|\phi\right\rangle^{(5)}_{1234}}{4}\left[a\left|01\right\rangle_{56}+b\left|00\right\rangle_{56}-c\left|11\right\rangle_{56}-d\left|10\right\rangle_{56}\right]
\nonumber \\
& + & \frac{\left|\phi\right\rangle^{(6)}_{1234}}{4}\left[a\left|01\right\rangle_{56}+b\left|00\right\rangle_{56}+c\left|11\right\rangle_{56}+d\left|10\right\rangle_{56}\right]
\nonumber \\
& + & \frac{\left|\phi\right\rangle^{(7)}_{1234}}{4}\left[-a\left|11\right\rangle_{56}+b\left|10\right\rangle_{56}+c\left|01\right\rangle_{56}-d\left|00\right\rangle_{56}\right]
\nonumber \\
& + & \frac{\left|\phi\right\rangle^{(8)}_{1234}}{4}\left[-a\left|11\right\rangle_{56}+b\left|10\right\rangle_{56}-c\left|01\right\rangle_{56}+d\left|00\right\rangle_{56}\right]
\nonumber \\
& + & \frac{\left|\phi\right\rangle^{(9)}_{1234}}{4}\left[a\left|10\right\rangle_{56}+b\left|11\right\rangle_{56}+c\left|00\right\rangle_{56}+d\left|01\right\rangle_{56}\right]
\nonumber \\
& + & \frac{\left|\phi\right\rangle^{(10)}_{1234}}{4}\left[-a\left|10\right\rangle_{56}-b\left|11\right\rangle_{56}+c\left|00\right\rangle_{56}+d\left|01\right\rangle_{56}\right]
\nonumber \\
& + & \frac{\left|\phi\right\rangle^{(11)}_{1234}}{4}\left[a\left|00\right\rangle_{56}-b\left|01\right\rangle_{56}+c\left|10\right\rangle_{56}-d\left|11\right\rangle_{56}\right]
\nonumber \\
& + & \frac{\left|\phi\right\rangle^{(12)}_{1234}}{4}\left[-a\left|00\right\rangle_{56}+b\left|01\right\rangle_{56}+c\left|10\right\rangle_{56}-d\left|11\right\rangle_{56}\right]
\nonumber \\
& + & \frac{\left|\phi\right\rangle^{(13)}_{1234}}{4}\left[-a\left|11\right\rangle_{56}-b\left|10\right\rangle_{56}+c\left|01\right\rangle_{56}+d\left|00\right\rangle_{56}\right]
\nonumber \\
& + & \frac{\left|\phi\right\rangle^{(14)}_{1234}}{4}\left[a\left|11\right\rangle_{56}+b\left|10\right\rangle_{56}+c\left|01\right\rangle_{56}+d\left|00\right\rangle_{56}\right]
\nonumber \\
& + & \frac{\left|\phi\right\rangle^{(15)}_{1234}}{4}\left[a\left|01\right\rangle_{56}-b\left|00\right\rangle_{56}-c\left|11\right\rangle_{56}+d\left|10\right\rangle_{56}\right]
\nonumber \\
& + & \frac{\left|\phi\right\rangle^{(16)}_{1234}}{4}\left[-a\left|01\right\rangle_{56}+b\left|00\right\rangle_{56}-c\left|11\right\rangle_{56}+d\left|10\right\rangle_{56}\right].
\end{eqnarray}

Thus if Alice makes a joint measurement on her particles ${(1234)}$, 
Bob's two particles (56) will be projected onto one of the sixteen equally
probable states. Bob recovers the information by applying
appropriate unitary transformations having Alice inform
him about her classical outcome(s). The advantage here is the
direct product of two single qubit unitary transformation listed in Table VII
instead of a joint unitary transformation such as a C-NOT gate for Bob to recover the unknown
information.

\subsection{\label{sec:level2}Second set}

In this subsection we propose another basis which is a linear
combination of direct products of the three-particle GHZ states and a single
particle as 
%
\begin{eqnarray}
\left| \varphi \right\rangle_{1234}^{(1),(2)}  = \frac{\left| \chi \rangle_{123}^{(1)''} \right. \otimes
\left| 0 \rangle_4 \right. \pm \left| \chi \rangle_{123}^{(3)''} \right. \otimes
\left| 1 \rangle_4 \right.  }{\sqrt{2}}, &  &
\left| \varphi \right\rangle_{1234}^{(3),(4)} = \frac{\left| \chi \rangle_{123}^{(2)''} \right. \otimes
\left| 0 \rangle_4 \right. \pm \left| \chi \rangle_{123}^{(4)''} \right. \otimes
\left| 1 \rangle_4 \right. }{\sqrt{2}}\ , \nonumber \\
\left| \varphi \right\rangle_{1234}^{(5),(6)}  = \frac{\left| \chi \rangle_{123}^{(1)''} \right. \otimes
\left| 1 \rangle_4 \right. \pm \left| \chi \rangle_{123}^{(3)''} \right. \otimes
\left| 0 \rangle_4 \right.  }{\sqrt{2}},  & &
\left| \varphi \right\rangle_{1234}^{(7),(8)} = \frac{\left| \chi \rangle_{123}^{(2)''} \right. \otimes
\left| 1 \rangle_4 \right. \pm \left| \chi \rangle_{123}^{(4)''} \right. \otimes
\left| 0 \rangle_4 \right. }{\sqrt{2}}\ , \nonumber \\
\left| \varphi \right\rangle_{1234}^{(9),(10)}  = \frac{\left| \chi \rangle_{123}^{(5)''} \right. \otimes
\left| 0 \rangle_4 \right. \pm \left| \chi \rangle_{123}^{(7)''} \right. \otimes
\left| 1 \rangle_4 \right.  }{\sqrt{2}} , &  &
\left| \varphi \right\rangle_{1234}^{(11),(12)} = \frac{\left| \chi \rangle_{123}^{(6)''} \right. \otimes
\left| 0 \rangle_4 \right. \pm \left| \chi \rangle_{123}^{(8)''} \right. \otimes
\left| 1 \rangle_4 \right. }{\sqrt{2}}\ , \nonumber \\
\left| \varphi \right\rangle_{1234}^{(13),(14)}  = \frac{\left| \chi \rangle_{123}^{(5)''} \right. \otimes
\left| 1 \rangle_4 \right. \pm \left| \chi \rangle_{123}^{(7)''} \right. \otimes
\left| 0 \rangle_4 \right.  }{\sqrt{2}}  & {~ ~ ~ \rm and} &
\left| \varphi \right\rangle_{1234}^{(15),(16)} = \frac{\left| \chi \rangle_{123}^{(6)''} \right. \otimes
\left| 1 \rangle_4 \right. \pm \left| \chi \rangle_{123}^{(8)''} \right. \otimes
\left| 0 \rangle_4 \right. }{\sqrt{2}} \nonumber \\ & & 
\end{eqnarray}

where
\begin{eqnarray}
\left| \chi \right\rangle_{123}^{(1)'',(2)''}  = \frac{1}{\sqrt{2}} \left[ \,
\left| 000 \right\rangle_{123} \pm \left| 111 \right\rangle_{123} \,
\right] & , &
\left| \chi \right\rangle_{123}^{(3)'',(4)''} = \frac{1}{\sqrt{2}} \left[ \,
\left| 010 \right\rangle_{123} \pm \left| 101 \right\rangle_{123} \,
\right] , \  \nonumber \\
\left| \chi \right\rangle_{123}^{(5)'',(6)''}  =\frac{1}{\sqrt{2}} \left[ \,
\left| 011 \right\rangle_{123} \pm \left| 100 \right\rangle_{123} \,
\right] & {\rm and} &  \left| \chi \right\rangle_{123}^{(7)'',(8)''} = \frac{1}{\sqrt{2}} \left[ \,
\left|001 \right\rangle_{123} \pm \left| 110 \right\rangle_{123} \,
\right]
\end{eqnarray}
are the eight GHZ states corresponding to three particles (123). \par
The following representation of the above basis set enables us to
generate a 
$2N$-particle entangled basis as described further, 
%
\begin{eqnarray}
\left| \chi \right\rangle_{1234}^{(1),(2)}  & = & \frac{\left| 0 \rangle_1
  \right. \otimes \left| \phi \rangle_{24}^+ \right. \otimes
\left| 0 \rangle_3 \right. \pm \left| 1 \rangle_1
  \right. \otimes \left| \psi \rangle_{24}^+ \right. \otimes
\left| 1 \rangle_3 \right.  }{\sqrt{2}} , \nonumber \\ 
\left| \chi \right\rangle_{1234}^{(3),(4)}  & = & \frac{\left| 0 \rangle_1
  \right. \otimes \left| \phi \rangle_{24}^+ \right. \otimes
\left| 1 \rangle_3 \right. \pm \left| 1 \rangle_1
  \right. \otimes \left| \psi \rangle_{24}^+ \right. \otimes
\left| 0 \rangle_3 \right.  }{\sqrt{2}} , \nonumber \\
\left| \chi \right\rangle_{1234}^{(5),(6)}   & = & \frac{\left| 0 \rangle_1
  \right. \otimes \left| \phi \rangle_{24}^- \right. \otimes
\left| 0 \rangle_3 \right. \pm \left| 1 \rangle_1
  \right. \otimes \left| \psi \rangle_{24}^- \right. \otimes
\left| 1 \rangle_3 \right.  }{\sqrt{2}} , \nonumber \\
\left| \chi \right\rangle_{1234}^{(7),(8)}  & = & \frac{\left| 0 \rangle_1
  \right. \otimes \left| \phi \rangle_{24}^- \right. \otimes
\left| 1 \rangle_3 \right. \pm \left| 1 \rangle_1
  \right. \otimes \left| \psi \rangle_{24}^- \right. \otimes
\left| 0 \rangle_3 \right.  }{\sqrt{2}} , \nonumber \\
\left| \chi \right\rangle_{1234}^{(9),(10)}  & = & \frac{\left| 1 \rangle_1
  \right. \otimes \left| \phi \rangle_{24}^+ \right. \otimes
\left| 0 \rangle_3 \right. \pm \left| 0 \rangle_1
  \right. \otimes \left| \psi \rangle_{24}^+ \right. \otimes
\left| 1 \rangle_3 \right.  }{\sqrt{2}}, \nonumber \\
\left| \chi \right\rangle_{1234}^{(11),(12)}  & = & \frac{\left| 1 \rangle_1
  \right. \otimes \left| \phi \rangle_{24}^+ \right. \otimes
\left| 1 \rangle_3 \right. \pm \left| 0 \rangle_1
  \right. \otimes \left| \psi \rangle_{24}^+ \right. \otimes
\left| 0 \rangle_3 \right.  }{\sqrt{2}}, \nonumber \\
\left| \chi \right\rangle_{1234}^{(13),(14)}  & = & \frac{\left| 1 \rangle_1
  \right. \otimes \left| \phi \rangle_{24}^- \right. \otimes
\left| 0 \rangle_3 \right. \pm \left| 0 \rangle_1
  \right. \otimes \left| \psi \rangle_{24}^- \right. \otimes
\left| 1 \rangle_3 \right.  }{\sqrt{2}} {~ ~ ~ \rm and} \nonumber \\
\left| \chi \right\rangle_{1234}^{(15),(16)}  & = & \frac{\left| 1 \rangle_1
  \right. \otimes \left| \phi \rangle_{24}^- \right. \otimes
\left| 1 \rangle_3 \right. \pm \left| 0 \rangle_1
  \right. \otimes \left| \psi \rangle_{24}^- \right. \otimes
\left| 0 \rangle_3 \right.  }{\sqrt{2}}. 
\end{eqnarray}

%
These set of states (Eq. 25/Eq. 27) are  also maximally entangled
four-particle states and cannot be written as direct products of 
lesser number of particles. Also the four-particle
correlation coefficients listed in Table VIII for the above states 
are non-zero and are also maximal ${(\pm 1)}$. The set has an additional advantage in
terms of robustness, i.e. when traced with respect to the 2nd or 4th
particle the remaining three particles (123, 134) are entangled. 
In addition, it is robust with respect to two-particle (13)
tracing which leaves other two particle (24) in an entangled state. This makes
the above set suitable for the teleportation of an arbitrary two-particle
state.  Using one of the above states as the state shared by
Alice and Bob, namely, 
\begin{equation}
\left| \varphi\right\rangle_{3456}^{(1)} =
\frac{1}{2}\left[\,\left|0000\right\rangle_{3456}+ \left|
    1110\right\rangle_{3456}+\left| 0101\right\rangle_{3456}+\left|
    1011\right\rangle_{3456} \right]
\end{equation}
where particles (34) are with Alice and particles (56) are with Bob, the
joint state of six-particles composed of Alice's particles
(1234) and Bob's particles  (56) give rise to
\begin{eqnarray}
\left| \psi \right\rangle_{123456} & = & \frac{a}{2}
\left[\left|
    000000 \right\rangle _{123456} +  \left|
    000101 \right\rangle _{123456} + \left|
    001110 \right\rangle _{123456} +  \left|
    001011 \right\rangle _{123456}  \right] \nonumber \\ & + & \frac{b}{2}\left[ \left|
    010000 \right\rangle _{123456} +  \left|
    010101 \right\rangle _{123456} + \left|
    011110 \right\rangle _{123456} +  \left|
    011011 \right\rangle _{123456}  \right] \nonumber \\ & + &
\frac{c}{2}
\left[\left|
    100000 \right\rangle _{123456} +  \left|
    100101 \right\rangle _{123456} + \left|
    101110 \right\rangle _{123456} +  \left|
    101011 \right\rangle _{123456}  \right] \nonumber \\ & +  & \frac{d}{2}\left[ \left|
    110000 \right\rangle _{123456} +  \left|
    110101 \right\rangle _{123456} + \left|
    111110 \right\rangle _{123456} +  \left|
    111011 \right\rangle _{123456}  \right].
\end{eqnarray}
Re-expressing  Eq. 29 in terms  of basis set proposed in (Eq. 25), we
have
\begin{eqnarray}
\left|\psi\right\rangle_{123456} & = &
\frac{\left|\varphi\right\rangle^{(1)}_{1234}}{4}\left[a\left|00\right\rangle_{56}+b\left|01\right\rangle_{56}+c\left|10\right\rangle_{56}+d\left|11\right\rangle_{56}\right]
\nonumber \\
& + & \frac{\left|\varphi\right\rangle^{(2)}_{1234}}{4}\left[a\left|00\right\rangle_{56}-b\left|01\right\rangle_{56}-c\left|10\right\rangle_{56}+d\left|11\right\rangle_{56}\right]
\nonumber \\
& + & \frac{\left|\varphi\right\rangle^{(3)}_{1234}}{4}\left[a\left|00\right\rangle_{56}+b\left|01\right\rangle_{56}-c\left|10\right\rangle_{56}-d\left|11\right\rangle_{56}\right]
\nonumber \\
& + & \frac{\left|\varphi\right\rangle^{(4)}_{1234}}{4}\left[a\left|00\right\rangle_{56}-b\left|01\right\rangle_{56}+c\left|10\right\rangle_{56}-d\left|11\right\rangle_{56}\right]
\nonumber \\
& + & \frac{\left|\varphi\right\rangle^{(5)}_{1234}}{4}\left[a\left|01\right\rangle_{56}+b\left|00\right\rangle_{56}+c\left|11\right\rangle_{56}+d\left|10\right\rangle_{56}\right]
\nonumber \\
& + & \frac{\left|\varphi\right\rangle^{(6)}_{1234}}{4}\left[a\left|01\right\rangle_{56}-b\left|00\right\rangle_{56}-c\left|11\right\rangle_{56}+d\left|10\right\rangle_{56}\right]
\nonumber \\
& + & \frac{\left|\varphi\right\rangle^{(7)}_{1234}}{4}\left[a\left|01\right\rangle_{56}+b\left|00\right\rangle_{56}-c\left|11\right\rangle_{56}-d\left|10\right\rangle_{56}\right]
\nonumber \\
& + & \frac{\left|\varphi\right\rangle^{(8)}_{1234}}{4}\left[a\left|01\right\rangle_{56}-b\left|00\right\rangle_{56}+c\left|11\right\rangle_{56}-d\left|10\right\rangle_{56}\right]
\nonumber \\
& + & \frac{\left|\varphi\right\rangle^{(9)}_{1234}}{4}\left[a\left|10\right\rangle_{56}+b\left|11\right\rangle_{56}+c\left|00\right\rangle_{56}+d\left|01\right\rangle_{56}\right]
\nonumber \\
& + & \frac{\left|\varphi\right\rangle^{(10)}_{1234}}{4}\left[-a\left|10\right\rangle_{56}+b\left|11\right\rangle_{56}+c\left|00\right\rangle_{56}-d\left|01\right\rangle_{56}\right]
\nonumber \\
& + & \frac{\left|\varphi\right\rangle^{(11)}_{1234}}{4}\left[a\left|10\right\rangle_{56}+b\left|11\right\rangle_{56}-c\left|00\right\rangle_{56}-d\left|01\right\rangle_{56}\right]
\nonumber \\
& + & \frac{\left|\varphi\right\rangle^{(12)}_{1234}}{4}\left[-a\left|10\right\rangle_{56}+b\left|11\right\rangle_{56}-c\left|00\right\rangle_{56}+d\left|01\right\rangle_{56}\right]
\nonumber \\
& + & \frac{\left|\varphi\right\rangle^{(13)}_{1234}}{4}\left[a\left|11\right\rangle_{56}+b\left|10\right\rangle_{56}+c\left|01\right\rangle_{56}+d\left|00\right\rangle_{56}\right]
\nonumber \\
& + & \frac{\left|\varphi\right\rangle^{(14)}_{1234}}{4}\left[-a\left|11\right\rangle_{56}+b\left|10\right\rangle_{56}+c\left|01\right\rangle_{56}-d\left|00\right\rangle_{56}\right]
\nonumber \\
& + & \frac{\left|\varphi\right\rangle^{(15)}_{1234}}{4}\left[a\left|11\right\rangle_{56}+b\left|10\right\rangle_{56}-c\left|01\right\rangle_{56}-d\left|00\right\rangle_{56}\right]
\nonumber \\
& + &
\frac{\left|\varphi\right\rangle^{(16)}_{1234}}{4}\left[-a\left|11\right\rangle_{56}+b\left|10\right\rangle_{56}-c\left|01\right\rangle_{56}+d\left|00\right\rangle_{56}\right]
.
\end{eqnarray}

It is obvious that, at the most, the unitary transformations which Bob needs to
apply reduce to a joint single qubit unitary transformation. Table
IX lists all the unitary transformations which might be needed
to recover the original state. A different orthogonal set of states
is given by
%
\begin{eqnarray}
\left| \varphi \right\rangle_{1234}^{(1)',(2)'}  = \frac{\left| \chi \rangle_{123}^{(1)''} \right. \otimes
\left| 0 \rangle_4 \right. \pm \left| \chi \rangle_{123}^{(4)''} \right. \otimes
\left| 1 \rangle_4 \right.  }{\sqrt{2}} & , &
\left| \varphi \right\rangle_{1234}^{(3)',(4)'} = \frac{\left| \chi \rangle_{123}^{(3)''} \right. \otimes
\left| 1 \rangle_4 \right. \pm \left| \chi \rangle_{123}^{(2)''} \right. \otimes
\left| 0 \rangle_4 \right. }{\sqrt{2}}\ , \nonumber \\
\left| \varphi \right\rangle_{1234}^{(5)',(6)'}  = \frac{\left| \chi \rangle_{123}^{(1)''} \right. \otimes
\left| 1 \rangle_4 \right. \pm \left| \chi \rangle_{123}^{(4)''} \right. \otimes
\left| 0 \rangle_4 \right.  }{\sqrt{2}} & , &
\left| \varphi \right\rangle_{1234}^{(7)',(8)'} = \frac{\left| \chi \rangle_{123}^{(3)''} \right. \otimes
\left| 0 \rangle_4 \right. \pm \left| \chi \rangle_{123}^{(2)''} \right. \otimes
\left| 1 \rangle_4 \right. }{\sqrt{2}}\ , \nonumber \\
\left| \varphi \right\rangle_{1234}^{(9)',(10)'}  = \frac{\left| \chi \rangle_{123}^{(5)''} \right. \otimes
\left| 0 \rangle_4 \right. \pm \left| \chi \rangle_{123}^{(8)''} \right. \otimes
\left| 1 \rangle_4 \right.  }{\sqrt{2}} & , &
\left| \varphi \right\rangle_{1234}^{(11)',(12)'} = \frac{\left| \chi \rangle_{123}^{(7)''} \right. \otimes
\left| 1 \rangle_4 \right. \pm \left| \chi \rangle_{123}^{(6)''} \right. \otimes
\left| 0 \rangle_4 \right. }{\sqrt{2}}\ , \nonumber \\
\left| \varphi \right\rangle_{1234}^{(13)',(14)'}  = \frac{\left| \chi \rangle_{123}^{(5)''} \right. \otimes
\left| 1 \rangle_4 \right. \pm \left| \chi \rangle_{123}^{(8)''} \right. \otimes
\left| 0 \rangle_4 \right.  }{\sqrt{2}}  & {\rm and} &
\left| \varphi \right\rangle_{1234}^{(15)',(16)'} = \frac{\left| \chi \rangle_{123}^{(7)''} \right. \otimes
\left| 0 \rangle_4 \right. \pm \left| \chi \rangle_{123}^{(6)''} \right. \otimes
\left| 1 \rangle_4 \right. }{\sqrt{2}} . \nonumber \\ & &
\end{eqnarray}
This set also has properties similar to the states given by Eq. 25/Eq. 27. The non-zero four-particle correlation
coefficient associated with all the sixteen basis states are listed in
Table X. It is an easy
exercise to verify that this set works successfully towards
teleportation of a two-particle system with only single qubit unitary
transformations on Bob's side. \par
It is possible to 
generalize the above protocol for the $N$-particle system using the basis
set given below. The sequential manner in which they are constructed
ensures that entanglement properties of these states are preserved
down to a pair of particles when they are systematically averaged. \par
 Consider the two-particle
Bell states given by Eq. 2 with particles 1 and 2 replaced by 2 and 3. The sixteen 
four-particle states (1234) can then be given as
\[ \frac{1}{\sqrt{2}} \left[ \left( \begin{array}{c} \left| 0
      \right\rangle \\
\left| 1 \right\rangle \\ \end{array} \right) _1
\otimes \left( \begin{array}{c} \left|  \phi ^+ \right\rangle \\
    \left| \psi ^+ \right\rangle \\ \end{array} \right) _{23}
\otimes \left( \begin{array}{c} \left| 0 \right\rangle \\ \left| 1
    \right\rangle  \\ \end{array} \right) _4  \pm
\left( \begin{array}{c} \left| 1 \right\rangle \\ \left| 0
    \right\rangle \\ \end{array} \right) _1
\otimes \left( \begin{array}{c} \left| \phi ^- \right\rangle \\ \left|
      \psi ^- \right\rangle  \\ \end{array} \right) _{23}
\otimes \left( \begin{array}{c} \left| 1 \right\rangle  \\ \left| 0
    \right\rangle \\ \end{array} \right) _4 \right] . \]
Relabelling the sixteen states as 
\begin{equation}
\left( \begin{array}{l} 
\left| \chi ^{(1),(2)} \right\rangle _{1234} \\
\left| \chi ^{(3),(4)} \right\rangle _{1234} \\
\left| \chi ^{(5),(6)} \right\rangle _{1234} \\
\left| \chi ^{(7),(8)} \right\rangle _{1234} \\
\left| \chi ^{(9),(10)} \right\rangle _{1234} \\  
\left| \chi ^{(11),(12)} \right\rangle _{1234} \\
\left| \chi ^{(13),(14)} \right\rangle _{1234} \\ 
\left| \chi ^{(15),(16)} \right\rangle _{1234} \\ 
\end{array}  \right) 
= \frac{1}{\sqrt{2}} \left [ \left( \begin{array}{c} 
\left| 0 \right\rangle \left| \phi^+ \right\rangle \left| 0 \right\rangle \\
\left| 0 \right\rangle \left| \phi^+ \right\rangle \left| 1 \right\rangle \\
\left| 0 \right\rangle \left| \psi^+ \right\rangle \left| 0 \right\rangle \\
\left| 0 \right\rangle \left| \psi^+ \right\rangle \left| 1 \right\rangle \\
\left| 1 \right\rangle \left| \phi^+ \right\rangle \left| 0 \right\rangle \\
\left| 1 \right\rangle \left| \phi^+ \right\rangle \left| 1 \right\rangle \\
\left| 1 \right\rangle \left| \psi^+ \right\rangle \left| 0 \right\rangle \\
\left| 1 \right\rangle \left| \psi^+ \right\rangle \left| 1 \right\rangle \\ 
\end{array} \right) \pm
\left( \begin{array}{c} 
\left| 1 \right\rangle \left| \phi^- \right\rangle \left| 1 \right\rangle \\
\left| 1 \right\rangle \left| \phi^- \right\rangle \left| 0 \right\rangle \\
\left| 1 \right\rangle \left| \psi^- \right\rangle \left| 1 \right\rangle \\
\left| 1 \right\rangle \left| \psi^- \right\rangle \left| 0 \right\rangle \\
\left| 0 \right\rangle \left| \phi^- \right\rangle \left| 1 \right\rangle \\
\left| 0 \right\rangle \left| \phi^- \right\rangle \left| 0 \right\rangle \\
\left| 0 \right\rangle \left| \psi^- \right\rangle \left| 1 \right\rangle \\
\left| 0 \right\rangle \left| \psi^- \right\rangle \left| 0 \right\rangle \\ 
\end{array} \right) \right ] ,
\end{equation}
the six-particle generalized entangled states are given by,
\[ \frac{1}{\sqrt{2}} \left [\left( \begin{array}{c} 0 \\ 1 \\ \end{array} \right) _1 \otimes 
\left( \begin{array}{l}
\left| \chi^{(1)} \right\rangle    _{2345} \\
\left| \chi^{(3)} \right\rangle    _{2345} \\
\left| \chi^{(5)} \right\rangle    _{2345}  \\
\left| \chi^{(7)} \right\rangle    _{2345}  \\
\left| \chi^{(9)} \right\rangle    _{2345}  \\
\left| \chi^{(11)} \right\rangle   _{2345}  \\
\left| \chi^{(13)} \right\rangle   _{2345}  \\
\left| \chi^{(15)} \right\rangle   _{2345}  \\
\end{array} \right) \otimes \left( \begin{array}{c} 0 \\ 1 \\ \end{array} \right) _6
\pm
\left( \begin{array}{c} 1 \\ 0 \\ \end{array} \right) _1 \otimes 
\left( \begin{array}{l}
\left| \chi^{(2)} \right\rangle    _{2345}  \\
\left| \chi^{(4)} \right\rangle    _{2345}  \\
\left| \chi^{(6)} \right\rangle    _{2345}  \\
\left| \chi^{(8)}  \right\rangle   _{2345}  \\
\left| \chi^{(10)} \right\rangle   _{2345}  \\
\left| \chi^{(12)}  \right\rangle  _{2345}  \\
\left| \chi^{(14} \right\rangle    _{2345}  \\
\left| \chi^{(16)} \right\rangle   _{2345} \\
\end{array} \right) \otimes \left( \begin{array}{c} 1 \\ 0 \\
\end{array} \right) _6 \right] . \]
The $2N$-particle generalization of the above which contains a set of  maximally entangled 
states can be written down immediately as
\begin{eqnarray}& & \frac{1}{\sqrt{2}} \left[ \left( \begin{array}{c} 0 \\ 1 \\ \end{array} \right) _1 \otimes 
\left( \begin{array}{l}
\left| \chi^{(1)}   \right\rangle _{23 \ldots 2N-1}  \\
\left| \chi^{(3)}   \right\rangle _{23 \ldots 2N-1} \\
 \vdots   \\
\left| \chi^{(2^{2N-2}-3)} \right\rangle _{23 \ldots 2N-1} \\
\left| \chi^{(2^{2N-2}-1)} \right\rangle _{23 \ldots 2N-1}  \\
\end{array} \right) \otimes \left( \begin{array}{c} 0 \\ 1 \\
\end{array} \right) _{2N} \right. \nonumber \\ & & 
\pm \left.
\left( \begin{array}{c} 1 \\ 0 \\ \end{array} \right) _1 \otimes 
\left( \begin{array}{l}
\left| \chi^{(2)}   \right\rangle _{23 \ldots 2N-1} \\
\left| \chi^{(4)}   \right\rangle _{23 \ldots 2N-1}  \\
 \vdots  \\
\left| \chi^{(2^{2N-2}-2)} \right\rangle _{23 \ldots 2N-1}  \\
\left| \chi^{(2^{2N-2})} \right\rangle _{23 \ldots 2N-1} \\
\end{array} \right) \otimes \left( \begin{array}{c} 1 \\ 0 \\
\end{array} \right) _{2N} \right] . \nonumber \end{eqnarray}

The $2N$-particle generalized entangled state for the second set can
also be obtained in a similar way as above.

\section*{IV.TELEPORTATION USING QUANTUM GATES AND COMPUTATIONAL
  BASIS}
In this section we analyze the above schemes using quantum gates and
the three and four-qubit computational basis.

\subsection*{A. Teleportation of a single qubit through GHZ state}

Three qubit states can be prepared by using the appropriate quantum
network as given below with the gates required [34-38] on the circuit for
the creation of the GHZ state(s). If we give three inputs $\left| 0
\right\rangle _1$, $\left| 0 \right\rangle _2$ and $\left| 0
\right\rangle _3$ then the GHZ state \ghza \ is prepared as indicated
in Fig. 1. The quantum circuit required for Alice's unknown qubit to be teleported is shown
in Fig. 2. The input for the quantum circuit in Fig. 2 is
\[ \left| \psi \right\rangle^{(0)}_{1234} = [a \left| 0 \right\rangle _1 + b
\left| 1 \right\rangle _1] \otimes \frac{1}{\sqrt{2}}
\left[ \left| 000 \right\rangle _{234}  +  \left| 111
\right\rangle _{234} \right] . \]
Alice sends her qubits 1 and 3 through the C-NOT gate keeping 1 as the
control qubit and 3 as the target qubit and obtains
%
\begin{equation}
\left| \psi \right\rangle^{(1)}_{1234} = \frac{a}{\sqrt{2}}
\left[ \left| 0000 \right\rangle _{1234}  +  \left| 0111
    \right\rangle _{1234} \right] + \frac{b}{\sqrt{2}}
\left[ \left| 1010 \right\rangle _{1234}  +  \left| 1101
    \right\rangle _{1234} \right] .
\end{equation}
%
Then she sends her qubit 1 through a Hadamard gate with the result
%
\begin{eqnarray}
\left| \psi \right\rangle ^{(2)}_{1234} &  = &  \frac{a}{2}
\left[ \left| 0000 \right\rangle _{1234}  +  \left| 1000
\right\rangle _{1234} +  \left| 0111 \right\rangle _{1234}  +  \left|
  1111 \right\rangle _{1234} \right] \nonumber \\ & + &
\frac{b}{2}
\left[ \left| 0010 \right\rangle _{1234}  - \left| 1010
\right\rangle _{1234} +  \left| 0101 \right\rangle _{1234}  - \left| 1101
\right\rangle _{1234} \right].
\end{eqnarray}
%
This she follows by sending qubit 2 through a Hadamard gate again which
results in the state as
%
\begin{eqnarray}
\left| \psi \right\rangle ^{(3)}_{1234} &  = &  \frac{a}{2\sqrt{2}}
\left[ \left| 0000 \right\rangle _{1234}  +  \left| 0100
\right\rangle _{1234} +  \left| 1000 \right\rangle _{1234}  +  \left|
  1100
\right\rangle _{1234} \right. \nonumber \\ & & + \left.
\left| 0011 \right\rangle _{1234}  -  \left| 0111
\right\rangle _{1234} +  \left| 1011 \right\rangle _{1234}  - \left| 1111
\right\rangle _{1234}  \right] \nonumber \\ & + &
\frac{b}{2\sqrt{2}}
 \left[ \left| 0010 \right\rangle _{1234}  +  \left| 0110
\right\rangle _{1234} -  \left| 1010 \right\rangle _{1234}  -  \left| 1110
\right\rangle _{1234} \right.  \nonumber \\  &  & + \left.
 \left| 0001 \right\rangle _{1234}  -  \left| 0101
\right\rangle _{1234} -  \left| 1001 \right\rangle _{1234}  + \left| 1101
\right\rangle _{1234}  \right].
\end{eqnarray}
%
A simple rearrangement will decompose the above state into four
equally probable measurement outcomes with Bob's particle being
projected in one of the four states as
\begin{eqnarray}
\left| \psi \right\rangle^{(3)}_{1234} & = & \frac{1}{2\sqrt{2}} \left\{
  \left| 000 \right\rangle _{123} \left[
    a \left| 0 \right\rangle _4 +b \left| 1 \right\rangle _4 \right] +
\left| 001 \right\rangle _{123}\left[ a \left| 1 \right\rangle _4 +b \left| 0 \right\rangle _4 \right] \right. \nonumber \\ & + &
\left| 010 \right\rangle _{123}\left[ a \left| 0 \right\rangle _4 -b \left| 1 \right\rangle _4 \right] +
\left| 011 \right\rangle _{123}\left[ -a \left| 1 \right\rangle _4 +b \left| 0 \right\rangle _4 \right]  \nonumber \\ & + &
\left| 110 \right\rangle _{123}\left[ a \left| 0 \right\rangle _4 +b \left| 1 \right\rangle _4 \right] +
\left| 101 \right\rangle _{123}\left[ a \left| 1 \right\rangle _4 -b \left| 0 \right\rangle _4 \right] \nonumber \\ & + &
\left.
  \left| 100 \right\rangle _{123} \left[ a \left| 0 \right\rangle _4 -b \left| 1 \right\rangle _4 \right] +
\left| 111 \right\rangle _{123}\left[ -a \left| 1 \right\rangle _4 -b \left| 0 \right\rangle _4 \right] \right..
\end{eqnarray}
Table XI lists the required gates on Bob's side which will be activated
based on Alice's communication through a classical channel.\par

\subsection*{B. Teleportation of an arbitrary EPR pair through GHZ basis}

The quantum circuit to accomplish this task is given in Figure 3.
In the figure, U's are single qubit unitary transformations on qubits
4 and 5 respectively. Qubits 1, 2 and 3 are with Alice and 4 and 5
are with Bob. The input to the quantum circuit is
%
\begin{eqnarray}
\left| \psi \right\rangle ^{(0)}_{12345} & = &  [a \left| 01 \right\rangle _{12} + b
\left| 10 \right\rangle _{12}] \otimes \frac{1}{\sqrt{2}}
\left[ \left| 000 \right\rangle _{345}  +  \left| 111
    \right\rangle _{345} \right] \nonumber \\ & = &
\frac{a}{\sqrt{2}} \left[ \left| 01000 \right\rangle _{12345}  +  \left| 01111
 \right\rangle _{12345} \right] + \frac{b}{\sqrt{2}}
\left[ \left| 10000 \right\rangle _{12345}  +  \left| 10111
 \right\rangle _{12345} \right]  . \end{eqnarray}
%
The sequence is Alice's transmission of her qubits 1 and 3 through
C-NOT gate while keeping qubit 1 as control and 3 as target. This she
follows with the transmission of 1 and 2 through the Hadamard gate. The
processes are given in the same sequence by the wave functions $\left|
  \psi \right\rangle^{(1)}_{12345} $, $\left|   \psi \right\rangle^{(2)}_{12345} $ and $\left|
  \psi \right\rangle^{(3)}_{12345} $, where
%
\begin{eqnarray}
\left| \psi \right\rangle^{(1)}_{12345} & = & \frac{a}{\sqrt{2}} \left[ \left|
    01000 \right\rangle _{12345} + \left|
    01111 \right\rangle _{12345} \right] +\frac{b}{\sqrt{2}} \left[ \left|
    10100 \right\rangle _{12345} + \left|
    10011 \right\rangle _{12345} \right], \\
\left| \psi \right\rangle ^{(2)}_{12345} & = & \frac{a}{2} \left[ \left|
    01000 \right\rangle _{12345} +  \left|
    11000 \right\rangle _{12345} + \left|
    01111 \right\rangle _{12345} +  \left|
    11111 \right\rangle _{12345} \right] \nonumber \\ & + &
 \frac{b}{2} \left[ \left|
    00100 \right\rangle _{12345} -  \left|
    10100 \right\rangle _{12345} + \left|
    00011 \right\rangle _{12345} -  \left|
    10011 \right\rangle _{12345} \right],   \\
& {\rm and} &  \nonumber \\
\left| \psi \right\rangle^{(3)}_{12345} & = & \frac{a}{2\sqrt{2}}
\left[\left|
    00000 \right\rangle _{12345} -  \left|
    01000 \right\rangle _{12345} + \left|
    10000 \right\rangle _{12345} -  \left|
    11000 \right\rangle _{12345}  \right. \nonumber \\ & & +\left. \left|
    00111 \right\rangle _{12345} -  \left|
    01111 \right\rangle _{12345} + \left|
    10111 \right\rangle _{12345} -  \left|
    11111 \right\rangle _{12345}  \right] \nonumber \\ & + &
\frac{b}{2\sqrt{2}}
\left[\left|
    00100 \right\rangle _{12345} +  \left|
    01100 \right\rangle _{12345} - \left|
    10100 \right\rangle _{12345} -  \left|
    11100 \right\rangle _{12345}  \right. \nonumber \\ & & +\left. \left|
    00011 \right\rangle _{12345} +  \left|
    01011 \right\rangle _{12345} - \left|
    10011 \right\rangle _{12345} -  \left|
    11011 \right\rangle _{12345}  \right].
 \end{eqnarray}
%
Decomposing this in terms of the computational three qubit (123) basis
set, we get,
%
\begin{eqnarray}
\left| \psi \right\rangle ^{(3)}_{12345} & = & \frac{1}{2\sqrt{2}} \left\{
  \left| 000 \right\rangle _{123} \left[
    a \left| 00 \right\rangle _{45} +b \left| 11 \right\rangle _{45} \right]+
\left| 010 \right\rangle _{123}\left[ -a \left| 00 \right\rangle _{45}
  +b \left| 11\right\rangle _{45} \right] \right. \nonumber \\ & + &
\left| 100 \right\rangle _{123}\left[ a \left| 00 \right\rangle _{45}
  -b \left|11 \right\rangle _{45} \right] +
\left| 110 \right\rangle _{123}\left[ -a \left| 00 \right\rangle _{45}
  -b \left| 11 \right\rangle _{45} \right]  \nonumber \\ & + &
\left| 001 \right\rangle _{123}\left[ a \left| 11 \right\rangle _{45}
  +b \left| 00 \right\rangle _{45} \right] +
\left| 011 \right\rangle _{123}\left[ -a \left| 11 \right\rangle _{45}
  +b \left| 00 \right\rangle _{45} \right] \nonumber \\ & + &
\left.
\left| 101 \right\rangle _{123} \left[ a \left| 11 \right\rangle _{45}
-b \left| 00 \right\rangle _{45} \right] +
\left| 111 \right\rangle _{123}\left[ -a \left| 11 \right\rangle _{45}
-b \left| 00 \right\rangle _{45} \right] \right\}.
\end{eqnarray}
%
It is clear from the above that Bob's measurements are all equally
probable and require at the most one two-qubit gate
leading to four equal outcomes. Table XII  gives the two
qubit gates required for measurements.\par
%
\subsection*{C. Teleportation of a single qubit through entangled basis
  of three qubits}
The three-qubit entangled basis (Eq. 6) can be prepared by applying the
Hadamard gate on the first qubit of GHZ basis as
\[ \frac{1}{\sqrt{2}} \left[ \, \left| 000 \right\rangle _{123} +
 \left| 111 \right\rangle _{123} \right]  \stackrel{H^1}{\longrightarrow}
\frac{1}{2} \left[ \, \left| 000 \right\rangle _{123} +
\left| 100 \right\rangle _{123} +\left| 011 \right\rangle _{123} -
\left| 111 \right\rangle _{123} \right]  \]
The quantum circuit to prepare the above state is given in Fig. 4.
The quantum circuit required to teleport the single qubit through
the above three qubit  entangled state is given in Fig. 5. 
The input to the circuit is
\begin{eqnarray}
\left| \psi \right\rangle^{(0)}_{1234} & = & \frac{a}{2} \left[ \left|
    0000 \right\rangle _{1234} +  \left|
    0011 \right\rangle _{1234} + \left|
    0100 \right\rangle _{1234} -  \left|
    0111 \right\rangle _{1234} \right] \nonumber \\ & + &
 \frac{b}{2} \left[ \left|
    1000 \right\rangle _{1234} +  \left|
    1011 \right\rangle _{1234} + \left|
    1100 \right\rangle _{1234} -  \left|
    1111 \right\rangle _{1234} \right] .
\end{eqnarray}
%
The four-qubit direct product state can be decomposed into four
equally probable results (similar to Eq. 16 and Eq. 17) on Bob's qubit as

\begin{eqnarray}
\left| \psi \right\rangle^{(3)}_{1234} & = & \frac{1}{2\sqrt{2}} \left\{
  \left| 000 \right\rangle _{123} \left[
    a \left| 0 \right\rangle _4 +b \left| 1 \right\rangle _4 \right] +
\left| 001 \right\rangle _{123}\left[ a \left| 1 \right\rangle _4 +b \left| 0 \right\rangle _4 \right] \right. \nonumber \\ & + &
\left| 010 \right\rangle _{123}\left[ a \left| 0 \right\rangle _4 -b \left| 1 \right\rangle _4 \right] +
\left| 011 \right\rangle _{123}\left[ -a \left| 1 \right\rangle _4 +b \left| 0 \right\rangle _4 \right]  \nonumber \\ & + &
\left| 110 \right\rangle _{123}\left[ a \left| 0 \right\rangle _4 +b \left| 1 \right\rangle _4 \right] +
\left| 101 \right\rangle _{123}\left[ a \left| 1 \right\rangle _4 -b \left| 0 \right\rangle _4 \right] \nonumber \\ & + &
\left.
  \left| 100 \right\rangle _{123} \left[ a \left| 0 \right\rangle _4 -b \left| 1 \right\rangle _4 \right] +
\left| 111 \right\rangle _{123}\left[ -a \left| 1 \right\rangle _4 -b \left| 0 \right\rangle _4 \right] \right..
\end{eqnarray}

%
The gates required for detection are summarized in Table XI.

\subsection*{D. Quantum circuits for two-qubit teleportation} 
Here we suggest the quantum circuits for preparing
different four-qubit entangled states and the network required to
teleport arbitrary two qubits through these quantum carriers using
four-qubit computational bases. \par
The quantum circuit required to prepare set of orthonormal states
given by Eq. 21 and Eq. 25 are given in figure 6 and figure 8
respectively. Depending on the input given all the 16 orthonormal
states can be prepared in the two different sets. In addition to this,
figures 7 and 9 provide the quantum network to teleport an
arbitrary two-qubit state using the four-particle entangled states (quantum carriers)
given by figures 6 and 8, respectively. The symbols have their usual meanings as
discussed earlier and algebra related to the process is
straightforward. The quantum network to
prepare six-qubit entangled state and to teleport a three qubit
arbitrary state through it can be developed on similar grounds. The
unitary transformation required on Bob's side are all single qubit
unitary transformations
and are quite simple to achieve.

\section*{V. CONCLUSION}
We have  given in this paper schemes for the single-particle teleportation through three-particle
GHZ states using three-particle entangled basis set(s)
and vice versa. The use of GHZ states and sets of three-particle
basis set(s) as {\em quantum carriers}
and as a set of {\em projection basis} has been explored. Our protocol
obviates the earlier difficulties regarding missing basis elements of
projection basis and Alice does not need any assistance
(Charlie/Cliff) in the process of communication with Bob. In that sense
ours is direct teleportation and not a controlled one, in contrast to schemes
proposed earlier. We have discussed the entanglement of multipartite
states using statistical correlation coefficients and have 
proposed multiparticle entangled states which possess
genuine multiparticle entanglement. We have demonstrated the
teleportation of arbitrary two-particle states
using the states proposed as quantum carriers with the added advantage
of the requirement of only the direct products of single qubit unitary transformations on
Bob's side instead of a joint unitary transformation involving two or
more particles.
 We have taken a step forward to suggest the generalization of the protocol
for the teleportation of the $N$-particle state through a $2N$-particle genuinely
entangled quantum channel which can be formed by taking proper care to
ensure maximum genuine entanglement. In addition,  we have
analyzed and verified all the protocols discussed here through the use of quantum
gates with appropriate quantum circuits.
\section*{VI. ACKNOWLEDGMENT}
AK is grateful to IIT Madras for a graduate fellowship. MSK would like
to acknowledge his gratitude to his mentor, Professor Bryan Sanctuary,
McGill University, Montreal, Canada for introducing him to the subject
of quantum teleportation. This project is funded by the IIT Madras
research funds.

\newpage
\begin{center} List of Tables \end{center}

\begin{enumerate}
\item Non-zero correlation coefficients associated with four Bell
      states (Table I).
\item Non-zero correlation coefficients associated with GHZ states
      and set of states given by Eq. 1 (Table II).
\item Unitary transformations for teleportation of a single particle
      through GHZ state (Table III).
\item Unitary transformations for teleportation of an arbitrary EPR
      pair through GHZ state (Table IV). 
\item Non-zero correlation coefficients associated with 16
      four-particle states proposed by Yeo and Chua [Ref. 18] (Table V).
\item Non-zero correlation coefficients associated with set of states
      given by Eq. 21 (Table VI).
\item Unitary transformations for teleportation of two-particle
      arbitrary state using $\left|\phi\right\rangle^{(1)}_{3456}$ as quantum carrier
      (Table VII).
\item Non-zero correlation coefficients associated with set of states
      given by Eq. 25 (Table VIII).
\item Unitary transformations for teleportation of two-particle
      arbitrary state using $\left|\varphi\right\rangle_{3456}^{(1)}$ as quantum carrier
      (Table IX).
\item Non-zero correlation coefficients associated with set of states
      given by Eq. 31 (Table X).
\item Quantum gates for recovering the unknown information sent by
      Alice (Table XI).
\item Single qubit quantum gates for quantum circuit  in Fig. 3 (Table XII).
\end{enumerate}

\newpage
\begin{center} List of Figures \end{center}
\begin{enumerate}
\item Quantum circuit to prepare GHZ state.
\item Quantum network to teleport single qubit through GHZ state.
\item Quantum network to teleport an EPR pair through GHZ state.
\item Quantum circuit to prepare $\left|\varphi\right\rangle_{123}^{(1)}$.
\item Quantum network to teleport single qubit state through
      $\left|\varphi\right\rangle_{123}^{(1)}$.
\item Quantum circuit to prepare $\left|\phi\right\rangle^{(1)}_{1234}$.
\item Quantum network to teleport arbitrary two qubit state $\left|\phi\right\rangle_{12}$ through
      $\left|\phi\right\rangle_{3456}^{(1)}$.
\item Quantum circuit to prepare $\left|\varphi\right\rangle^{(1)}_{1234}$.
\item Quantum network to teleport arbitrary two qubit state $\left|\phi\right\rangle_{12}$ through
      $\left|\varphi\right\rangle_{3456}^{(1)}$.

\end{enumerate}

%
%
%
\newpage
\begin{center}
Table I
\end{center} 
\begin{center}
\begin{tabular}{|l|c|c|c|c|}  \hline
 & $\left| \psi \right\rangle_{12}^- $ & $\left| \psi \right\rangle_{12}^+ $ &
$\left| \phi \right\rangle_{12}^- $ & $\left| \phi
\right\rangle_{12}^+ $ \\ [2.0 ex] \hline
$C_{xx}^{12}$ & -1 & 1 & -1 & 1 \\  [2.0 ex]
$C_{yy}^{12}$ & -1 & 1 & 1 & -1 \\  [2.0 ex] 
$C_{zz}^{12}$ & -1 & -1 & 1 & 1 \\  [2.0 ex] \hline \end{tabular}
\end{center}

%

\newpage
\begin{center}
Table II
\end{center} 
\begin{center}
\begin{tabular}{|c|c|c|c|c|c|c|}  \hline
& $C_{xxx}^{123}$ & $C_{yyx}^{123}$ & $C_{yxy}^{123}$ & $C_{xyy}^{123}$ & $C_{xxz}^{123}$ &
$C_{yyz}^{123}$ \\ [2.0 ex] \hline
$\frac{1}{\sqrt{2}} \left[ \,\left| 000 \right\rangle_{123} \pm \left| 111
  \right\rangle_{123} \, \right]$ & $\pm 1$ & $\mp 1$ & $\mp 1$ & $\mp 1$ &
- & - \\  [2.0 ex]
$\frac{1}{\sqrt{2}} \left[ \,\left| 001 \right\rangle_{123} \pm \left| 110
  \right\rangle_{123} \, \right]$ & $\pm 1$ & $\mp 1$ & $\pm 1$ & $\pm 1$ &
- & - \\  [2.0 ex]
$\frac{1}{\sqrt{2}} \left[ \,\left| 010 \right\rangle_{123} \pm \left| 101
  \right\rangle_{123} \, \right]$ & $\pm 1$ & $\pm 1$ & $\mp 1$ & $\pm 1$ &
- & - \\  [2.0 ex] 
$\frac{1}{\sqrt{2}} \left[ \,\left| 011 \right\rangle_{123} \pm \left| 100
  \right\rangle_{123} \, \right]$ & $\pm 1$ & $\pm 1$ & $\pm 1$ & $\mp 1$ &
- & - \\  [2.0 ex] 
$\left| \chi \right\rangle^{(1)}_{123} $ & - & - &  1 &  1 &  1 & -1
\\  [2.0 ex] 
$\left| \chi \right\rangle^{(2)}_{123} $ & - & - & -1 & -1 &  1 & -1
\\  [2.0 ex] 
$\left| \chi \right\rangle^{(3)}_{123} $ & - & - & -1 & -1 & -1 &  1
\\  [2.0 ex] 

$\left| \chi \right\rangle^{(4)}_{123} $ & - & - &  1 &  1 & -1 &  1
\\  [2.0 ex] 

$\left| \chi \right\rangle^{(5)}_{123} $ & - & - &  1 & -1 &  1 &  1
\\  [2.0 ex] 
$\left| \chi \right\rangle^{(6)}_{123} $ & - & - & -1 &  1 &  1 &  1
\\  [2.0 ex] 

$\left| \chi \right\rangle^{(7)}_{123} $ & - & - & -1 &  1 & -1 & -1
\\   [2.0 ex] 

$\left| \chi \right\rangle^{(8)}_{123} $ & - & - &  1 & -1 & -1 & -1
\\ 
 [2.0 ex] \hline \end{tabular} \end{center}
%
%

\newpage
\begin{center}
Table III
\end{center} 
\begin{center}
\begin{tabular}{|c|c|}
\hline  Measurement & Unitary \\
outcome & transformation \\  [2.0 ex]  \hline
\ $ \left| \chi \right\rangle^{(1)}_{123}$ \ , \ $ \left| \chi
\right\rangle^{(4)}_{123}$ \ 
 & $\sigma^{4}_{z}$ \\  [2.0 ex] 
\ $ \left| \chi\right\rangle^{(2)}_{123}$ \ , \ $ \left| \chi
\right\rangle^{(3)}_{123}$ \ 
 & $I^{4}$ \\  [2.0 ex] 
\ $ \left| \chi \right\rangle^{(5)}_{123}$ \ , \ $ \left| \chi
\right\rangle^{(8)}_{123}$ \ 
& $\sigma^{4}_{x}$ \\  [2.0 ex] 
\ $ \left| \chi \right\rangle^{(6)}_{123}$ \ , $ \left| \chi
\right\rangle^{(7)}_{123}$ \ 
 & $\sigma^{4}_{y}$ \\  [2.0 ex] \hline  \end{tabular}
\end{center}

\newpage
\begin{center}
Table IV
\end{center} 
\begin{center}
\begin{tabular}{|c|c|}
\hline  Measurement & Unitary \\
outcome & transformation \\  [2.0 ex]  \hline
\ $ \left| \varphi \right\rangle ^{(1)}_{123}$ \ , \ $ \left| \varphi
\right\rangle^{(4)}_{123}$ \
 & $\sigma^{4}_{x} \otimes I^5$ \\  [2.0 ex] 
\ $ \left| \varphi \right\rangle^{(2)}_{123}$ \ , \ $ \left| \varphi
\right\rangle ^{(3)}_{123}$ \ 
 & $\sigma^{4}_{x} \otimes \sigma^{5}_{z}$ \\  [2.0 ex] 
\ $ \left| \varphi \right\rangle ^{(5)}_{123}$ \ , \ $ \left| \varphi
\right\rangle ^{(8)}_{123}$ \ 
& $I^4 \otimes \sigma^{5}_{x}$ \\  [2.0 ex] 
\ $ \left| \varphi \right\rangle ^{(6)}_{123}$ \ , \ $ \left| \varphi
\right\rangle ^{(7)}_{123}$ \ 
 & $\sigma^{4}_{z}\otimes \sigma^{5}_{x}$ \\  [2.0 ex]  \hline \end{tabular}
\end{center}
%

\newpage
\begin{center}
Table V
\end{center} 
\begin{eqnarray*}
\left| \chi^{00} \right\rangle _{1234} & = & \frac
{1}{\sqrt{2}}\left[\left|\zeta\right\rangle^{0}_{1234}+\left|\zeta\right\rangle^{1}_{1234}
  \right]  {~~~ \rm where}
\end{eqnarray*}
\begin{eqnarray*}
\left|\zeta\right\rangle^{0} & = & \frac
{1}{2}\left[\left|0000\right\rangle-\left|0011\right\rangle
  -\left|0101\right\rangle +\left|0110\right\rangle  \right] {~~~ \rm
  and} \, \, \, \left|\zeta\right\rangle^{1} = \frac
{1}{2}\left[\left|1001\right\rangle+\left|1010\right\rangle
  +\left|1100\right\rangle +\left|1111\right\rangle  \right]
\end{eqnarray*}
\begin{center}
\begin{tabular}{|c|c|c|c|c|}  \hline
& $C_{xyyx}^{1234}$ & $C_{xzzx}^{1234}$ & $C_{zyyz}^{1234}$ &
$C_{zzzz}^{1234}$ \\ [2.0 ex] \hline
$\left| \chi^{00} \right\rangle _{1234}$ & -1  & +1  & -1 & +1 \\ [2.0 ex]
$I^1 \otimes \sigma_{x}^2$ \,   $\left| \chi^{00} \right\rangle _{1234}$ & +1  & -1  & +1 & -1 \\
[2.0 ex]
$I^1 \otimes \sigma_{y}^2$ \, $\left| \chi^{00} \right\rangle _{1234}$ & -1  & -1  & -1 & -1 \\
[2.0 ex]
$I^1 \otimes \sigma_{z}^2$ \,   $\left| \chi^{00} \right\rangle _{1234}$ & +1  & +1  & +1 & +1 \\
[2.0 ex]
$\sigma_{x}^1 \otimes I^2$ \,   $\left| \chi^{00} \right\rangle _{1234}$ & -1  & +1  & +1 & -1 \\
[2.0 ex]
$\sigma_{x}^1 \otimes \sigma_{x}^2$ \,   $\left| \chi^{00} \right\rangle _{1234}$ & +1  & -1  & -1 & +1 \\
[2.0 ex]
$\sigma_{x}^1 \otimes \sigma_{y}^2$ \,   $\left| \chi^{00} \right\rangle _{1234}$ & -1  & -1  & +1 & +1 \\
[2.0 ex]
$\sigma_{x}^1 \otimes \sigma_{z}^2$ \,   $\left| \chi^{00} \right\rangle _{1234}$ & +1  & +1  & -1 & -1 \\
[2.0 ex]
$\sigma_{y}^1 \otimes I^2$ \,   $\left| \chi^{00} \right\rangle _{1234}$ & +1  & -1  & +1 & -1 \\
[2.0 ex]
$\sigma_{y}^1 \otimes \sigma_{x}^2$ \,   $\left| \chi^{00} \right\rangle _{1234}$ & -1  & +1  & -1 & +1 \\
[2.0 ex]
$\sigma_{y}^1 \otimes \sigma_{y}^2$ \,   $\left| \chi^{00} \right\rangle _{1234}$ & +1  & +1  & +1 & +1 \\
[2.0 ex]
$\sigma_{y}^1 \otimes \sigma_{z}^2$ \,   $\left| \chi^{00} \right\rangle _{1234}$ & -1  & -1  & -1 & -1 \\
[2.0 ex]
$\sigma_{z}^1 \otimes I^2$ \,   $\left| \chi^{00} \right\rangle _{1234}$ & +1  & -1  & -1 & +1 \\
[2.0 ex]
$\sigma_{z}^1 \otimes \sigma_{x}^2$ \,   $\left| \chi^{00} \right\rangle _{1234}$ & -1  & +1  & +1 & -1 \\
[2.0 ex]
$\sigma_{z}^1\otimes \sigma_{y}^2$ \,   $\left| \chi^{00} \right\rangle _{1234}$ & +1  & +1  & -1 & -1 \\
[2.0 ex]
$\sigma_{z}^1 \otimes \sigma_{z}^2$ \,   $\left| \chi^{00} \right\rangle _{1234}$ & -1  & -1  & +1 & +1 \\ [2.0 ex]
\hline \end{tabular} \end{center}

\newpage
\begin{center}
Table VI
\end{center} 
\begin{center}
\begin{tabular}{|c|c|c|c|c|}  \hline
& $C_{xxyy}^{1234}$ & $C_{xyxy}^{1234}$ & $C_{yxyx}^{1234}$ &
$C_{yyxx}^{1234}$ \\ [2.0 ex] \hline
$ \left| \phi \right\rangle ^{(1)}_{1234}$ & +1  & +1  & +1 & +1 \\ [2.0 ex]
$ \left| \phi \right\rangle ^{(2)}_{1234}$ & -1  & -1  & -1 & -1 \\
[2.0 ex]
$ \left| \phi \right\rangle ^{(3)}_{1234}$ & -1  & -1  & +1 & +1 \\
[2.0 ex]
$ \left| \phi \right\rangle ^{(4)}_{1234}$ & +1  & +1  & -1 & -1 \\
[2.0 ex]
$ \left| \phi \right\rangle ^{(5)}_{1234}$ & -1  & +1  & -1 & +1 \\
[2.0 ex]
$ \left| \phi \right\rangle ^{(6)}_{1234}$ & +1  & -1  & +1 & -1 \\
[2.0 ex]
$ \left| \phi \right\rangle ^{(7)}_{1234}$ & +1  & -1  & -1 & +1 \\
[2.0 ex]
$ \left| \phi \right\rangle ^{(8)}_{1234}$ & -1  & +1  & +1 & -1 \\
[2.0 ex]
$ \left| \phi \right\rangle ^{(9)}_{1234}$ & +1  & +1  & -1 & -1 \\
[2.0 ex]
$ \left| \phi \right\rangle ^{(10)}_{1234}$ & -1  & -1  & +1 & +1 \\
[2.0 ex]
$ \left| \phi \right\rangle ^{(11)}_{1234}$ & -1  & -1  & -1 & -1 \\
[2.0 ex]
$ \left| \phi \right\rangle ^{(12)}_{1234}$ & +1  & +1  & +1 & +1 \\
[2.0 ex]
$ \left| \phi \right\rangle ^{(13)}_{1234}$ & -1  & +1  & +1 & -1 \\
[2.0 ex]
$ \left| \phi \right\rangle ^{(14)}_{1234}$ & +1  & -1  & -1 & +1 \\
[2.0 ex]
$ \left| \phi \right\rangle ^{(15)}_{1234}$ & +1  & -1  & +1 & -1 \\
[2.0 ex]
$ \left| \phi \right\rangle ^{(16)}_{1234}$ & -1  & +1  & -1 & +1 \\ [2.0 ex]
\hline \end{tabular} \end{center}

\newpage
\begin{center}
Table VII
\end{center} 
\begin{center}
\begin{tabular}{|c|c|}
\hline  Measurement & Unitary \\
outcome & transformation \\  [2.0 ex]  \hline
\ $ \left| \phi \right\rangle ^{(1)}_{1234}$
 & $I^{5} \otimes I^{6}$ \\  [2.0 ex] 
\ $ \left| \phi \right\rangle ^{(2)}_{1234}$
& $\sigma^{5}_{z} \otimes I^{6}$ \\  [2.0 ex] 
\ $ \left| \phi \right\rangle ^{(3)}_{1234}$
 & $\sigma^{5}_{x} \otimes\sigma^{6}_{z}$ \\
 [2.0 ex] 

\ $ \left| \phi \right\rangle ^{(4)}_{1234}$
& $\sigma^{5}_{z} \otimes \sigma^{6}_{z} \otimes\sigma^{5}_{x}$ \\
 [2.0 ex]

\ $ \left| \phi \right\rangle ^{(5)}_{1234}$
  & $\sigma^{5}_{z} \otimes\sigma^{6}_{x}$ \\  [2.0 ex] 

\ $ \left| \phi \right\rangle ^{(6)}_{1234}$
 & $I^{5} \otimes\sigma^{6}_{x}$ \\
 [2.0 ex] 

\ $ \left| \phi \right\rangle ^{(7)}_{1234}$
& $\sigma^{5}_{z} \otimes \sigma^{6}_{z} \otimes\sigma^{5}_{x}  \otimes\sigma^{6}_{x}$ \\
 [2.0 ex]  
\ $ \left| \phi \right\rangle ^{(8)}_{1234}$
 & $\sigma^{5}_{x} \otimes \sigma^{6}_{x} \otimes\sigma^{6}_{z}$ \\
 [2.0 ex]

\ $ \left| \phi \right\rangle ^{(9)}_{1234}$
& $\sigma^{5}_{x} \otimes I^{6}$ \\
 [2.0 ex] 

\ $ \left| \phi \right\rangle ^{(10)}_{1234}$
 & $\sigma^{5}_{x} \otimes \sigma^{5}_{z} \otimes I^{6}$ \\
 [2.0 ex] 

\ $ \left| \phi \right\rangle ^{(11)}_{1234}$
& $I^{5} \otimes \sigma^{6}_{z}$ \\  [2.0 ex] 

\ $ \left| \phi \right\rangle ^{(12)}_{1234}$
& $\sigma^{5}_{z} \otimes\sigma^{6}_{z}$ \\  [2.0 ex]

\ $ \left| \phi \right\rangle ^{(13)}_{1234}$
& $\sigma^{5}_{x} \otimes \sigma^{6}_{x} \otimes\sigma^{5}_{z}$ \\
 [2.0 ex]  
\ $ \left| \phi \right\rangle ^{(14)}_{1234}$
& $\sigma^{5}_{x} \otimes \sigma^{6}_{x}$ \\
 [2.0 ex] 

\ $ \left| \phi \right\rangle ^{(15)}_{1234}$
 & $\sigma^{5}_{z} \otimes \sigma^{6}_{z} \otimes\sigma^{6}_{x}$ \\
 [2.0 ex] 
\ $ \left| \phi \right\rangle ^{(16)}_{1234}$
& $I^{5} \otimes \sigma^{6}_{z} \otimes\sigma^{6}_{x}$ \\
 [2.0 ex] 
 \hline  \end{tabular}
\end{center}

\newpage
\begin{center}
Table VIII
\end{center} 
\begin{center}
\begin{tabular}{|c|c|c|c|c|}  \hline
& $C_{xyyz}^{1234}$ & $C_{xzyy}^{1234}$ & $C_{yyxz}^{1234}$ &
$C_{yzxy}^{1234}$ \\ [2.0 ex] \hline
$ \left| \varphi \right\rangle ^{(1)}_{1234}$ & -1  & -1  & -1 & -1 \\ [2.0 ex]
$ \left| \varphi \right\rangle ^{(2)}_{1234}$ & -1  & +1  & -1 & +1 \\
[2.0 ex]
$ \left| \varphi \right\rangle ^{(3)}_{1234}$ & +1  & +1  & +1 & +1 \\
[2.0 ex]
$ \left| \varphi \right\rangle ^{(4)}_{1234}$ & +1  & -1  & +1 & -1 \\
[2.0 ex]
$ \left| \varphi \right\rangle ^{(5)}_{1234}$ & +1  & +1  & +1 & +1 \\
[2.0 ex]
$ \left| \varphi \right\rangle ^{(6)}_{1234}$ & +1  & -1  & +1 & -1 \\
[2.0 ex]
$ \left| \varphi \right\rangle ^{(7)}_{1234}$ & -1  & -1  & -1 & -1 \\
[2.0 ex]
$ \left| \varphi \right\rangle ^{(8)}_{1234}$ & -1  & +1  & -1 & +1 \\
[2.0 ex]
$ \left| \varphi \right\rangle ^{(9)}_{1234}$ & -1  & -1  & +1 & +1 \\
[2.0 ex]
$ \left| \varphi \right\rangle ^{(10)}_{1234}$ & -1  & +1  & +1 & -1 \\
[2.0 ex]
$ \left| \varphi \right\rangle ^{(11)}_{1234}$ & +1  & +1  & -1 & -1 \\
[2.0 ex]
$ \left| \varphi \right\rangle ^{(12)}_{1234}$ & +1  & -1  & -1 & +1 \\
[2.0 ex]
$ \left| \varphi \right\rangle ^{(13)}_{1234}$ & +1  & +1  & -1 & -1 \\
[2.0 ex]
$ \left| \varphi \right\rangle ^{(14)}_{1234}$ & +1  & -1  & -1 & +1 \\
[2.0 ex]
$ \left| \varphi \right\rangle ^{(15)}_{1234}$ & -1  & -1  & +1 & +1 \\
[2.0 ex]
$ \left| \varphi \right\rangle ^{(16)}_{1234}$ & -1  & +1  & +1 & -1 \\ [2.0 ex]
\hline \end{tabular} \end{center}

\newpage
\begin{center}
Table IX
\end{center} 
\begin{center}
\begin{tabular}{|c|c|}
\hline  Measurement & Unitary \\
outcome & transformation \\  [2.0 ex]  \hline
\ $ \left| \varphi \right\rangle ^{(1)}_{1234}$
 & $I^{5} \otimes I^{6}$ \\  [2.0 ex] 
\ $ \left| \varphi \right\rangle ^{(2)}_{1234}$
 & $ \sigma^{5}_{z} \otimes \sigma^{6}_{z}$ \\  [2.0 ex] 
\ $ \left| \varphi \right\rangle ^{(3)}_{1234}$
 & $\sigma_{z}^{5} \otimes I^{6}$ \\  [2.0 ex] 
\ $ \left| \varphi \right\rangle ^{(4)}_{1234}$
 & $I^{5} \otimes \sigma_{z}^{6}$ \\  [2.0 ex] 
\ $ \left| \varphi \right\rangle ^{(5)}_{1234}$
 & $I^{5} \otimes\sigma^{6}_{x}$ \\  [2.0 ex] 
\ $ \left| \varphi \right\rangle ^{(6)}_{1234}$
 & $\sigma^{5}_{z} \otimes \sigma^{6}_{z} \otimes\sigma^{6}_{x}$ \\
 [2.0 ex] 
\ $ \left| \varphi \right\rangle ^{(7)}_{1234}$
 & $\sigma^{6}_{x} \otimes\sigma^{5}_{z}$ \\
 [2.0 ex] 
\ $ \left| \varphi \right\rangle ^{(8)}_{1234}$
 & $I^{5} \otimes \sigma^{6}_{z} \otimes\sigma^{6}_{x}$ \\
 [2.0 ex] 
\ $ \left| \varphi \right\rangle ^{(9)}_{1234}$
 &$\sigma^{5}_{x} \otimes I^{6}$  \\
 [2.0 ex] 
\ $ \left| \varphi \right\rangle ^{(10)}_{1234}$
 &  $\sigma^{6}_{z} \otimes \sigma^{5}_{x} \otimes\sigma^{5}_{z}$ \\
 [2.0 ex] 
\ $ \left| \varphi \right\rangle ^{(11)}_{1234}$
 & $\sigma^{5}_{z} \otimes\sigma^{5}_{x} \otimes I^{6}$ \\
 [2.0 ex] 
\ $ \left| \varphi \right\rangle ^{(12)}_{1234}$
 & $ \sigma^{6}_{z} \otimes\sigma^{5}_{x} $  \\
 [2.0 ex] 
\ $ \left| \varphi \right\rangle ^{(13)}_{1234}$
 & $\sigma^{5}_{x} \otimes \sigma^{6}_{x}$  \\
 [2.0 ex] 
\ $ \left| \varphi \right\rangle ^{(14)}_{1234}$
 & $\sigma^{5}_{x} \otimes \sigma^{6}_{x} \otimes\sigma^{5}_{z}  \otimes\sigma^{6}_{z}$ \\
 [2.0 ex] 
\ $ \left| \varphi \right\rangle ^{(15)}_{1234}$
 & $\sigma^{5}_{z} \otimes \sigma^{5}_{x} \otimes\sigma^{6}_{x}$ \\
 [2.0 ex] 
\ $ \left| \varphi \right\rangle ^{(16)}_{1234}$
 & $\sigma^{6}_{z} \otimes \sigma^{6}_{x} \otimes\sigma^{5}_{x}$ \\
 [2.0 ex] \hline  \end{tabular}
\end{center}

\newpage
\begin{center}
Table X
\end{center} 
\begin{center}
\begin{tabular}{|c|c|c|c|c|}  \hline
& $C_{xxxz}^{1234}$ & $C_{xzxx}^{1234}$ & $C_{yxyz}^{1234}$ &
$C_{yzyx}^{1234}$ \\ [2.0 ex] \hline
$ \left| \varphi \right\rangle ^{(1)'}_{1234}$ & +1  & -1  & -1 & +1 \\ [2.0 ex]
$ \left| \varphi \right\rangle ^{(2)'}_{1234}$ & +1  & +1  & -1 & -1 \\
[2.0 ex]
$ \left| \varphi \right\rangle ^{(3)'}_{1234}$ & -1  & +1  & +1 & -1 \\
[2.0 ex]
$ \left| \varphi \right\rangle ^{(4)'}_{1234}$ & -1  & -1  & +1 & +1 \\
[2.0 ex]
$ \left| \varphi \right\rangle ^{(5)'}_{1234}$ & -1  & -1  & +1 & +1 \\
[2.0 ex]
$ \left| \varphi \right\rangle ^{(6)'}_{1234}$ & -1  & +1  & +1 & -1 \\
[2.0 ex]
$ \left| \varphi \right\rangle ^{(7)'}_{1234}$ & +1  & +1  & -1 & -1 \\
[2.0 ex]
$ \left| \varphi \right\rangle ^{(8)'}_{1234}$ & +1  & -1  & -1 & +1 \\
[2.0 ex]
$ \left| \varphi \right\rangle ^{(9)'}_{1234}$ & +1  & +1  & +1 & +1 \\
[2.0 ex]
$ \left| \varphi \right\rangle ^{(10)'}_{1234}$ & +1  & -1  & +1 & -1 \\
[2.0 ex]
$ \left| \varphi \right\rangle ^{(11)'}_{1234}$ & -1  & -1  & -1 & -1 \\
[2.0 ex]
 $ \left| \varphi \right\rangle ^{(12)'}_{1234}$ & -1  & +1  & -1 & +1 \\
[2.0 ex]
$ \left| \varphi \right\rangle ^{(13)'}_{1234}$ & -1  & +1  & -1 & +1 \\
[2.0 ex]
$ \left| \varphi \right\rangle ^{(14)'}_{1234}$ & -1  & -1  & -1 & -1 \\
[2.0 ex]
$ \left| \varphi \right\rangle ^{(15)'}_{1234}$ & +1  & -1  & +1 & -1 \\
[2.0 ex]
$ \left| \varphi \right\rangle ^{(16)'}_{1234}$ & +1  & +1  & +1 & +1 \\ [2.0 ex]
\hline \end{tabular} \end{center}

\newpage
\begin{center}
Table XI
\end{center} 
\begin{center}
\begin{tabular}{|c|c|}
\hline & \\  Measurement outcome & Unitary transformation \\ [1.0
ex]
\hline \ $\left| 000 \right\rangle _{123}$ \ and \ $\left|
  110\right\rangle_{123}$ \ & $I^4$ \\   [2.0 ex] 
\ $\left| 001 \right\rangle_{123}$ \ and   \ $\left|
  111\right\rangle_{123}$ \ &
$X$ gate \\  [2.0 ex] 
\ $\left| 010 \right\rangle _{123}$ \ and   \ $\left|
  100\right\rangle_{123}$ \ & $Z$ gate \\  [2.0 ex] 
\ $\left| 101 \right\rangle _{123}$ \ and   \ $\left| 011\right\rangle
_{123}$ \ & $X$ gate followed by $Z$ gate \\ [2.0 ex]  \hline
\end{tabular} \end{center}
%
%

\newpage
\begin{center}
Table XII
\end{center} 
\begin{center}
\begin{tabular}{|c|c|}
\hline & \\  Measurement outcome & Unitary transformation \\  [2.0 ex] 
\hline \ $\left| 000 \right\rangle _{123}$ \ and   \ $\left| 110
\right\rangle _{123}$ \
  & $I^4 \otimes X^5$ gate \\   [1.5 ex]  \ $\left| 001 \right\rangle
  _{123}$  \ and
\ $\left| 111 \right\rangle _{123}$ \ &  $X^4 \, \, {\rm gate} \otimes I^5$  \ \\  [2.0 ex] 
\ $\left| 010 \right\rangle _{123}$ \ and   \ $\left| 100 \right\rangle
_{123}$ \
&  $Z^4 \, \, {\rm gate} \otimes X^5 \, \, {\rm gate}$   \\  [2.0 ex]  \ $\left| 101 \right\rangle
_{123}$ \ and  \ $\left| 011 \right\rangle _{123}$ \ &  $X^4 \, \, {\rm gate} \otimes Z^5 \, \, {\rm gate}$ 
\\  [2.0 ex]  \hline \end{tabular} \end{center}

Note : superscript on a prescribed quantum gates ($X$ or $Z$) represents the particle index
to which the quantum gate must be applied.

%
%
\newpage
\begin{center}
Figure 1
\end{center}
\begin{figure}[!htb]
\vspace*{0.5in}
\includegraphics [width=8.5in] {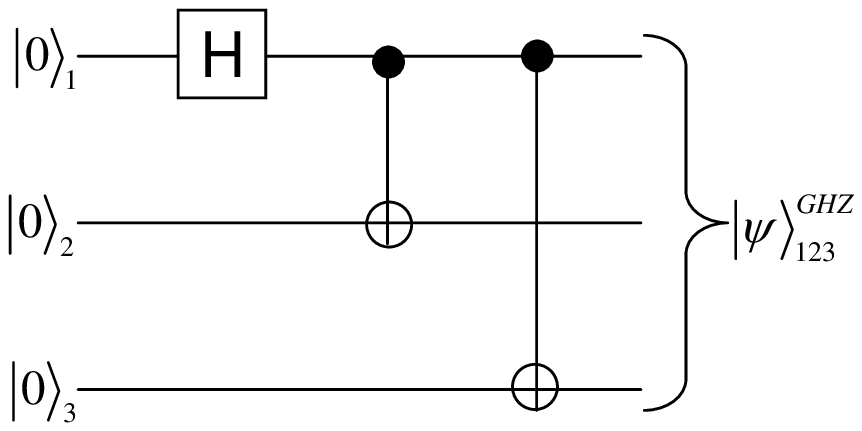}
\vspace*{-1.0in}
\end{figure}

\newpage
\begin{center}
Figure 2
\end{center}
\begin{figure}[!htb]
\vspace*{0.5 in}
\includegraphics [width=7.5in] {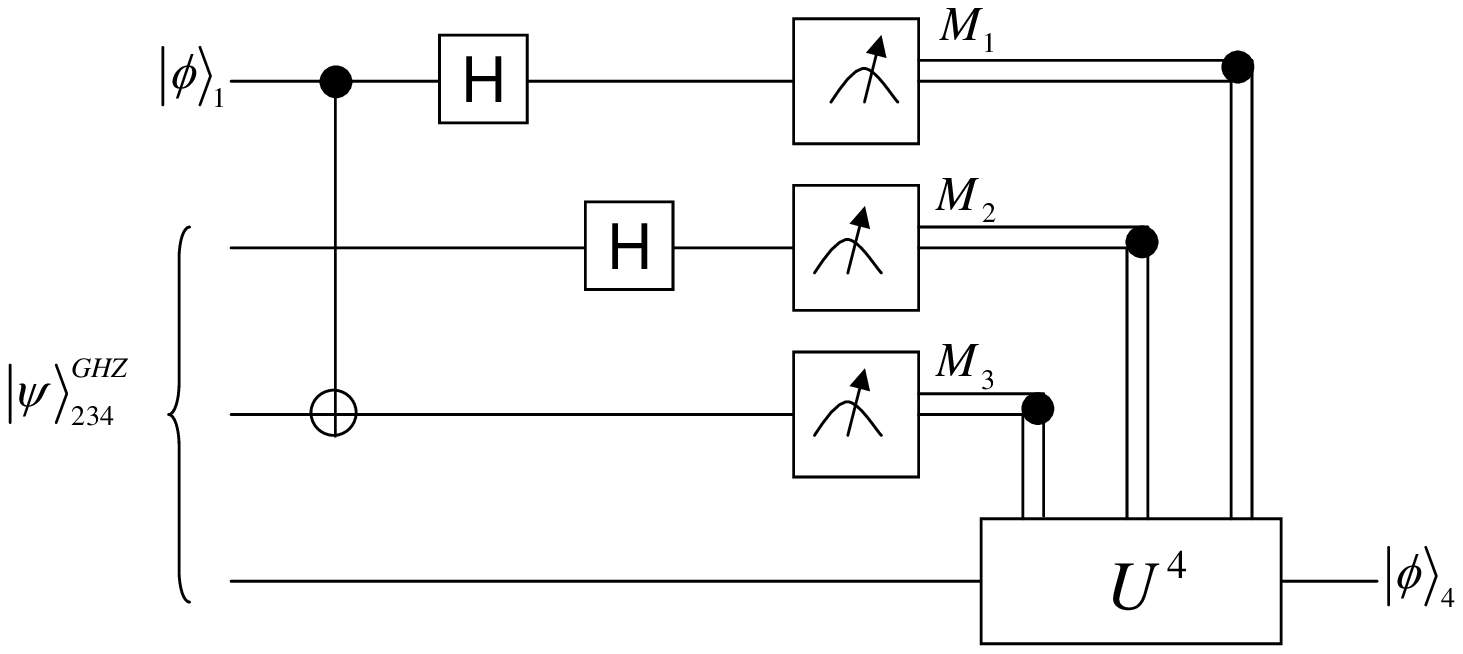}
\vspace*{-1.0 in}
\end{figure}   

\newpage
\begin{center}
Figure 3
\end{center}
\begin{figure}[!htb]
\vspace*{0.5 in}
\includegraphics [width=8.5in] {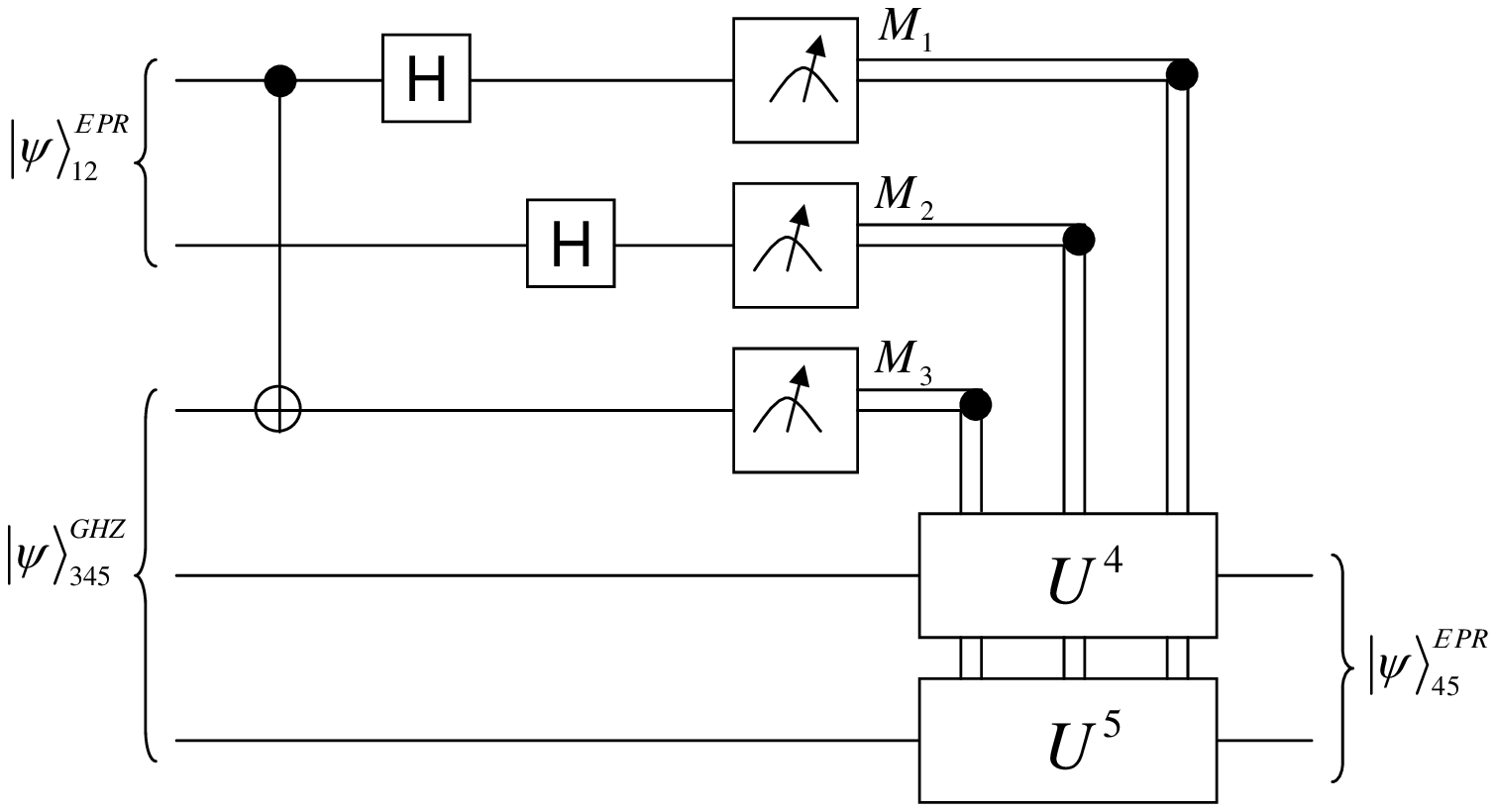}
\vspace*{-1.0 in}
\end{figure}

\newpage
\begin{center}
Figure 4
\end{center}
\begin{figure}[!htb]
\vspace*{0.5 in}
\includegraphics [width=8.5in] {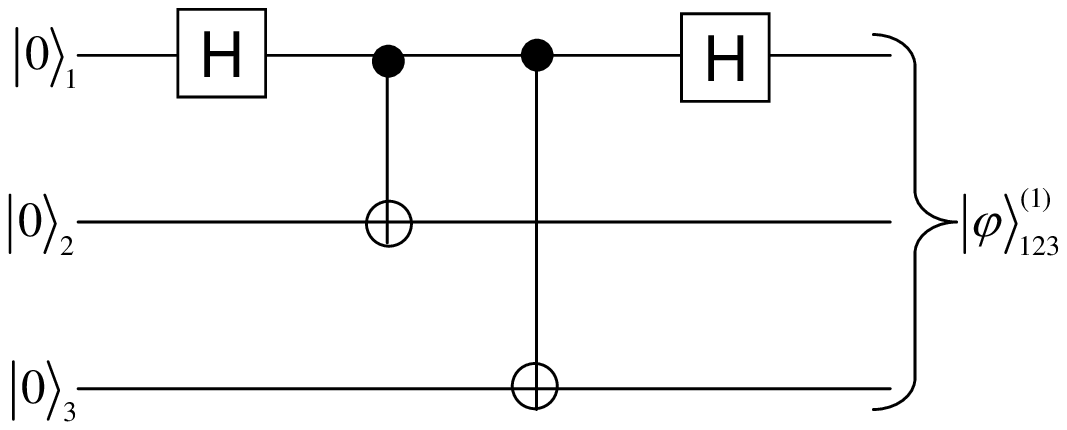}
\vspace*{-1.0 in}
\end{figure} 

\newpage
\begin{center}
Figure 5
\end{center}
\begin{figure}[!htb]
\vspace*{0.5 in}
\includegraphics [width=8.5in] {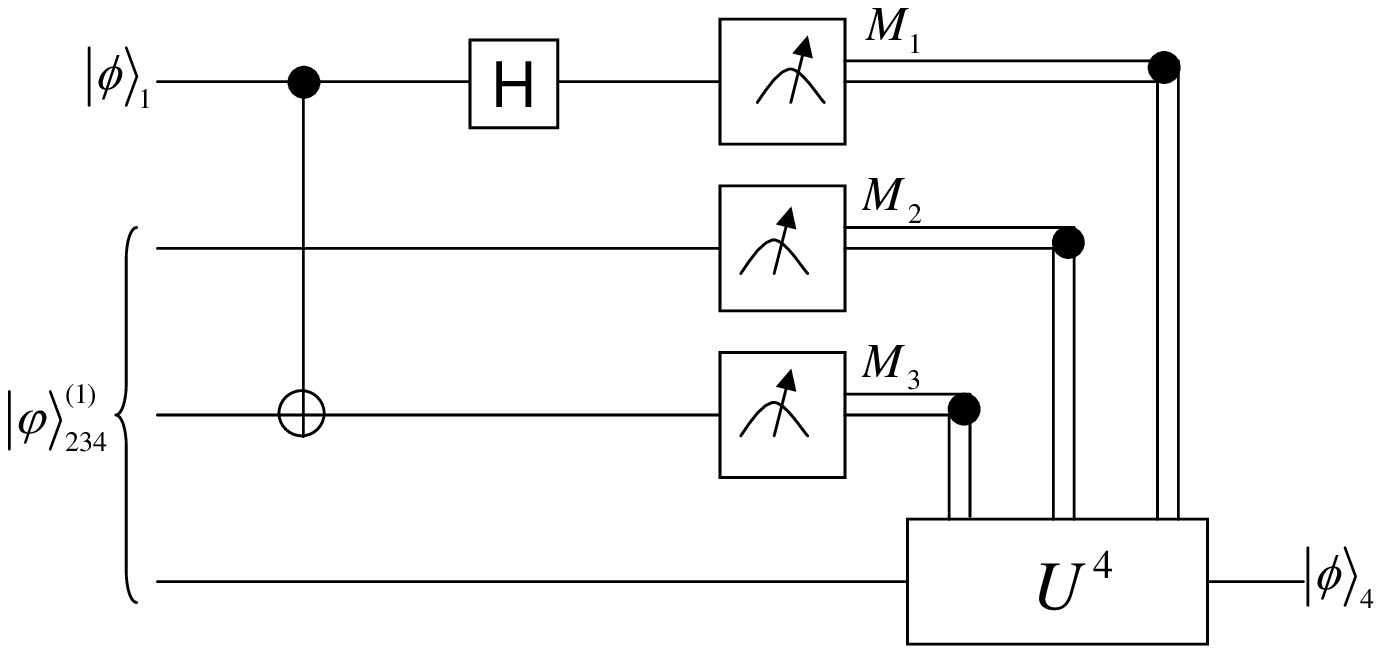}
\vspace*{-1.0 in}
\end{figure}

\newpage
\begin{center}
Figure 6
\end{center}
\begin{figure}[!htb]
\vspace*{0.5 in}
\includegraphics [width=8.5in] {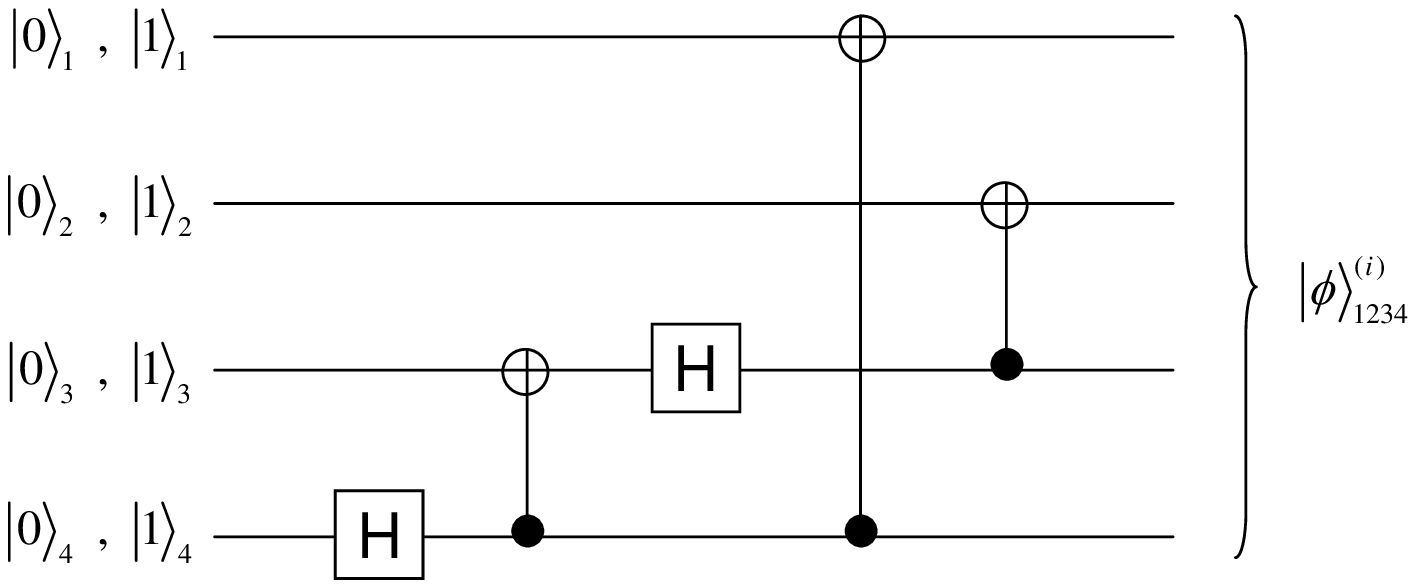}
\vspace*{-1.0 in}
\end{figure}

\newpage
\begin{center}
Figure 7
\end{center}
\begin{figure}[!htb]
\vspace*{0.5 in}
\includegraphics [width=8.5in] {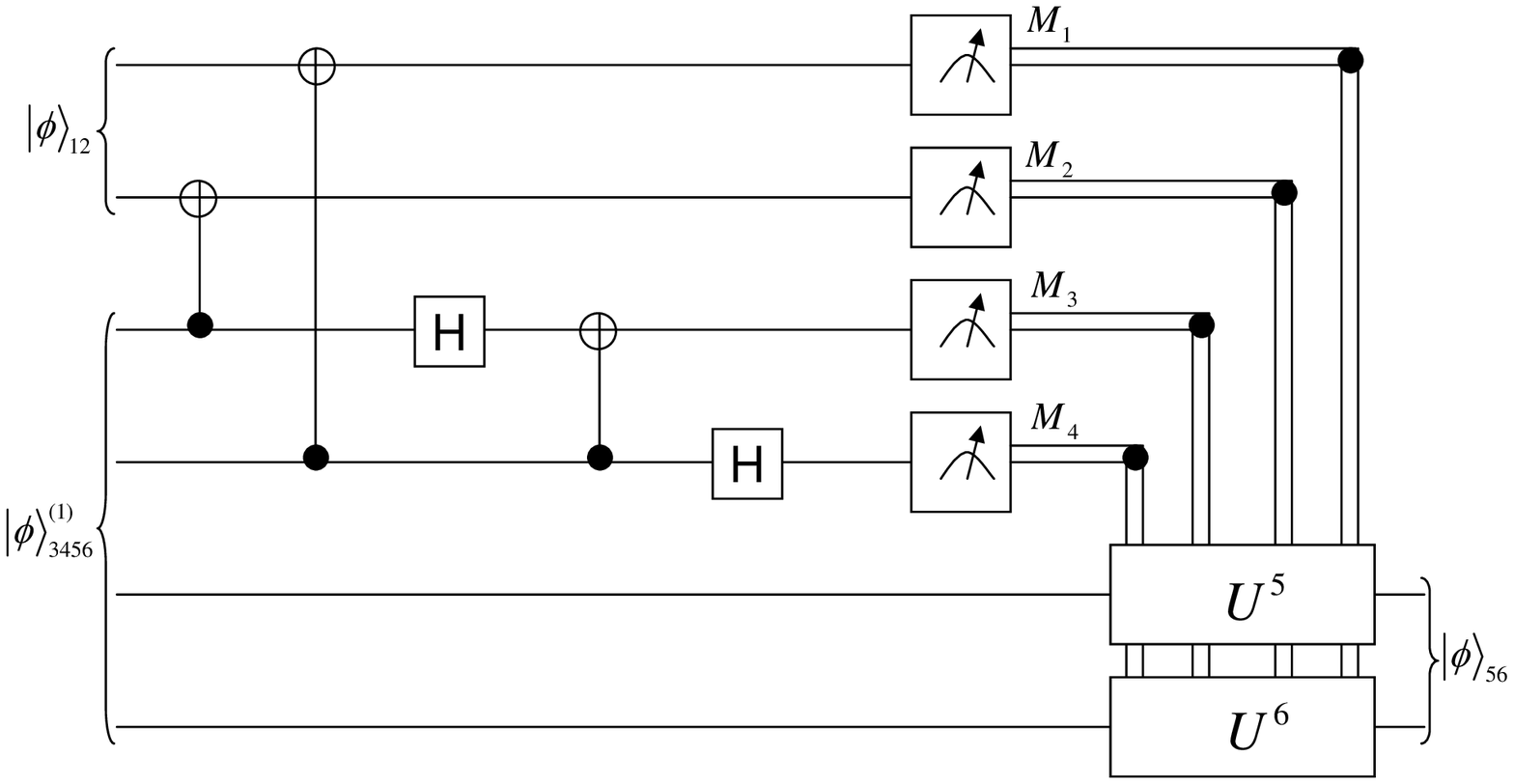}
\vspace*{-1.0 in}
\end{figure}

\newpage
\begin{center}
Figure 8
\end{center}
\begin{figure}[!htb]
\vspace*{0.5 in}
\includegraphics [width=8.5in] {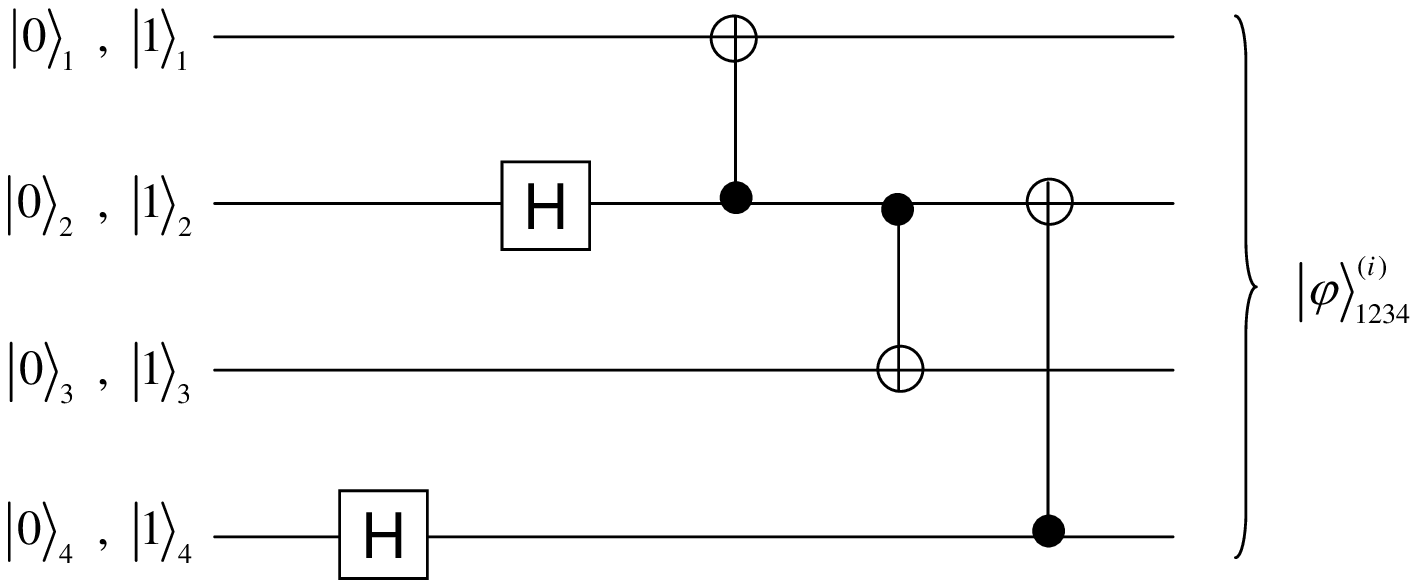}
\vspace*{-1.0 in}
\end{figure}

\newpage
\begin{center}
Figure 9
\end{center}
\begin{figure}[!htb]
\vspace*{0.5 in}
\includegraphics [width=8.5in] {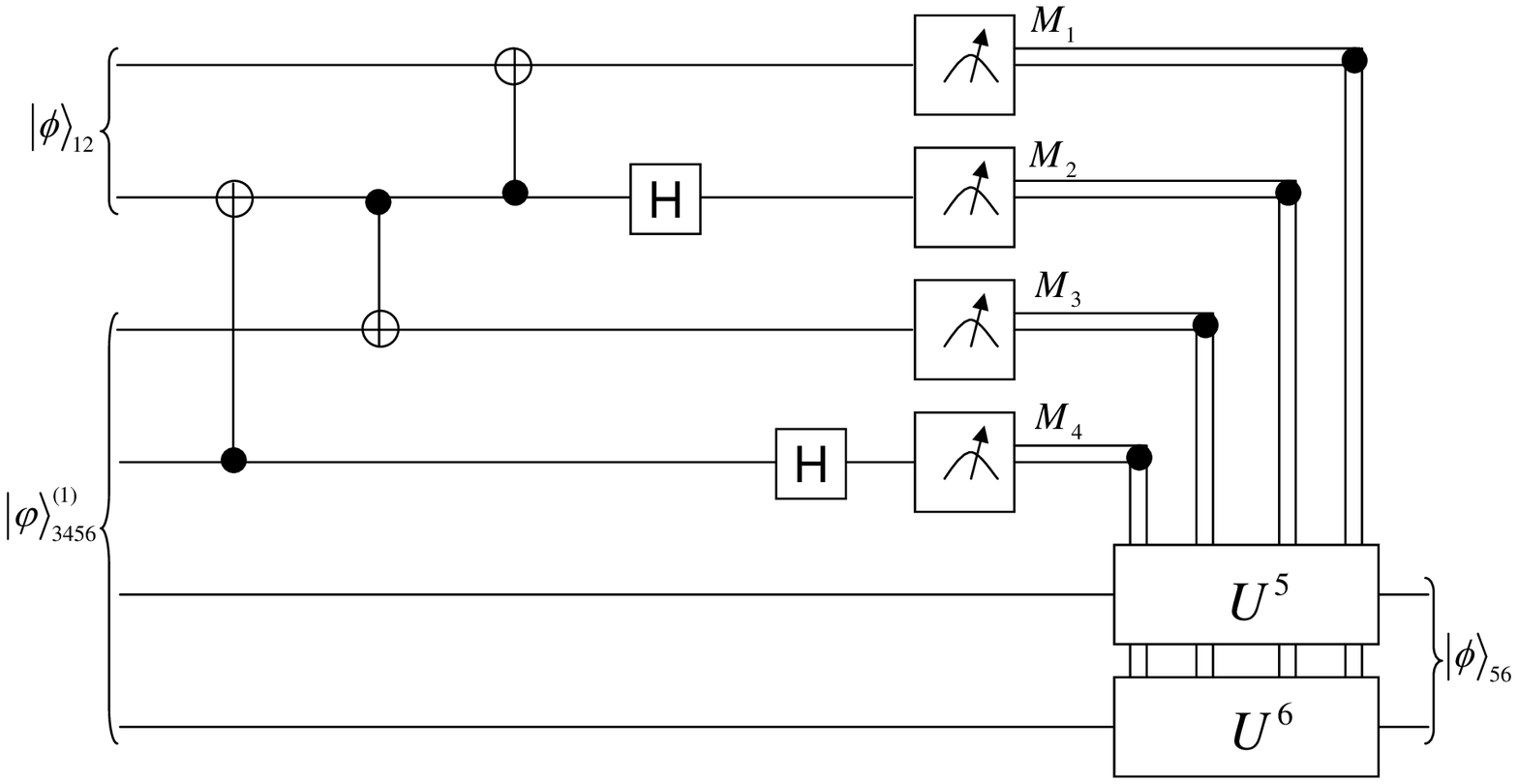}
\vspace*{-1.0 in}
\end{figure}


\newpage
\begin{thebibliography}{50}
\bibitem{Bennett1}  C. H.  Bennett, G. Brassard, C. Crepeau, R. Jozsa,
        A. Peres and, W. K. Wootters, Phys. Rev. Lett. {\bf70}, 1895 (1993).
\bibitem{BoAhar}  D. Bohm and Y.  Aharanov, Phys. Rev.  {\bf108}, 1070 (1957).
\bibitem{EPR}  A. Einstein, B. Podolsky and, N. Rosen, Phys. Rev.
      {\bf47}, 777 (1935).
\bibitem{nature} J. F. Sherson, H. Krauter, R. K. Olsson,
        B. Julsgaard, K. Hammerer, I. Cirac and, E. S. Polzik, Nature
        {\bf443}, 557 (2006).
\bibitem{Boumees1}  D. Bouwmeester, J. W. Pan, K. Mattle, M. Eibl,
        H. Weinfurter and, A. Zeilinger, Nature  {\bf390}, 575 (1997).
\bibitem{brauns1} S. L. Braunstein and
        H. J.  Kimble, Nature  {\bf394}, 840 (1998).
\bibitem{furasawa1}  A. Furusawa, J. L. Sorensen, S. L. Braunstein, C.
        A. Fuchs, H. J. Kimble and, E. S. Polzik, Science  {\bf282}, 706 (1998).
\bibitem{boschi}  D. Boschi, S. Branca, F. De Martini, L. Hardy and, S.
        Popescu, Phys. Rev. Lett.  {\bf80}, 1121 (1998).
\bibitem{kim1}  Y. H. Kim, S. P. Kulik and, Y. Shih, Phys. Rev. Lett.  {\bf86}, 1370 (2001).
\bibitem{boumees2}  D. Bouwmeester,  K. Mattle, J. W. Pan,
        H. Weinfurter, A. Zeilinger and, M. Zukowski, Applied Physics B
        {\bf67}, 749 (1998).
\bibitem{karlsson}  A. Karlsson and M. Bourennane, Phys. Rev. A  {\bf58}, 4394 (1998).
\bibitem{shi1}  B. S. Shi and A. Tomita, Phys. Lett. A  {\bf296}, 161 (2002).
\bibitem{gorbachev}  V. N. Gorbachev and A. I. Trubilko, JETP  {\bf91}, 894 (2000).
\bibitem{rigolin}  G. Rigolin, Phys. Rev. A  {\bf71}, 032303 (2005).
\bibitem{fang}  J. Fang, Y. Lin, S. Zhu and, X. Chen, Phys. Rev. A  {\bf67}, 014305 (2003).
\bibitem{deng1}  F. G. Deng, C. Y. Li, Y. S. Li, H. Y. Zhou and, Y. Wang, Phys. Rev. A  {\bf72}, 022338 (2005).
\bibitem{fengli} Y. Feng-Li and D. He-Wei, Chin. Phys. Lett.  {\bf23},
      17 (2006).
\bibitem{Yeo} Y. Yeo and, W. K. Chua, Phys. Rev. Lett. {\bf96}, 060502
              (2006).
\bibitem{boumees3}  D. M. Greenberg, M. Horne and, A. Zeilinger, {\em Bell's
        Theorem, Quantum Theory, and Conceptions of the Universe}, edited
        by M. Kafatos (Kluwer, Dordrecht, 1989); D. Bouwmeester,
        J. W. Pan, M. Daniell, H. Weinfurter and, A. Zeilinger, Phys. Rev. Lett.  {\bf82}, 1345 (1999).
\bibitem{pan2}  J. W. Pan, M. Daniell, S. Gasparoni, G. Weihs and,
        A. Zeilinger. Phys. Rev. Lett. {\bf86}, 4435 (2001).
\bibitem{weinfurter}  H. Weinfurter and M. Zukowski, Phys. Rev. A  {\bf64}, 010102 (2001).
\bibitem{zhao}  Z. Zhao, Y. A. Chen, A. N. Zhang, H. J. Briegel and,
        J. W. Pan, Nature  {\bf430}, 54 (2004).
\bibitem{bennett2}  C. H. Bennett, H. J. Bernstein, S. Popescu and, B. Schumacher, Phys. Rev. A  {\bf53}, 2046 (1996).
\bibitem{wooters1}  S. Hill and W. K. Wootters, Phys. Rev. Lett.  {\bf78}, 5022 (1997).
\bibitem{wooters2}  W. K. Wootters, Phys. Rev. Lett.  {\bf80}, 2245 (1998).

\bibitem{schlienz}  J. Schlienz and G. Mahler, Phys. Rev. A  {\bf52}, 4396 (1995).
\bibitem{altafini} C. Altafini, Phys. Rev. A  {\bf69}, 012311 (2004).
\bibitem{huang}  K. Huang, {\em Statistical Mechanics} (John Wiley, New
         York, 1987).
\bibitem{osborn}  T. A. Osborn, Phys. Rev. A  {\bf16}, 334 (1977).
\bibitem{Watson}  K. M. Watson, Phys. Rev.  {\bf103}, 489 (1956).
\bibitem{dur1}  W. Dur, G. Vidal and, J. I. Cirac, Phys. Rev. A  {\bf62}, 062314 (2000).
\bibitem{dur2}  W. Dur, Phys. Rev. A  {\bf63}, 020303 (2001).
\bibitem{rajagopal}  A. K. Rajagopal and R. W. Rendell, Phys. Rev. A  {\bf65}, 032328 (2002).
\bibitem{Barenco} A. Barenco, C. H. Bennett, R. Cleve,
        D. P. DiVincenzo, N. Margolus, P. Shor, T. Sleator,
        J. A. Smolin and, H. Weinfurter,  Phys. Rev. A  {\bf52}, 3457 (1995).
\bibitem{Zhou} X. Zhou, D. W. Leung and, I. L. Chuang, Phys. Rev. A  {\bf62}, 052316 (2000).
\bibitem{Brassard} G. Brassard, S. L. Braunstein and, R. Cleve, Physica
        D  {\bf120}, 43 (1998).
\bibitem{Vedral} V. Vedral and M. B. Plenio, Progress in Quantum
        Electronics  {\bf22}, 1 (1998).
\bibitem{Nielsen} M. A. Nielsen and I. L. Chuang, {\em Quantum
        Computation and Quantum Information} (Cambridge University
        Press, 2000).
\end{thebibliography}
\end{document}